\RequirePackage{fix-cm}
\documentclass[referee]{svjour3}
\smartqed  
\usepackage{amssymb}
\usepackage{fullpage}
\usepackage{comment}
\usepackage{color}
\usepackage{subfigure}
\usepackage{amsmath}
\usepackage{graphicx}
\usepackage{verbatim}
\usepackage[ansinew]{inputenc}
\usepackage{epsfig}
\usepackage{verbatim}
\usepackage{amsfonts}
\usepackage{setspace}
\usepackage{soul}
\usepackage{multirow}
\usepackage{wrapfig}
\usepackage[lined,boxed,commentsnumbered,ruled]{algorithm2e}
\usepackage[usenames,dvipsnames]{xcolor}
\usepackage{program}
\usepackage{algpseudocode}
\usepackage[ansinew]{inputenc}
\usepackage[hyperfootnotes=false]{hyperref}
\hypersetup{
  colorlinks   = true, 
  urlcolor     = blue, 
  linkcolor    = blue, 
  citecolor   = blue 
}
\usepackage{url}

\begin{document}

\title{Uni-MUMAC: A Unified Down/Up-link MU-MIMO MAC Protocol for IEEE 802.11ac WLANs
}


\author{Ruizhi Liao \and Boris Bellalta \and Trang Cao Minh \and \\ Jaume Barcelo \and Miquel Oliver}


\institute{R. Liao \and B. Bellalta \and T.C. Minh \and J. Barcelo \and M. Oliver \at
              Department of Information and Communication Technologies \\
	      Universitat Pompeu Fabra, Barcelona, 08018, Spain \\
              \email{\{ruizhi.liao, boris.bellalta, trang.cao, jaume.barcelo, miquel.oliver\}@upf.edu}           
}


\maketitle

\begin{abstract}
Due to the dominance of the downlink traffic in Wireless Local Area Networks (WLANs), a large number of previous research efforts have been put to enhance the transmission from the Access Point (AP) to stations (STAs). The downlink Multi-User Multiple-Input Multiple-Output (MU-MIMO) technique, supported by the latest IEEE amendment-802.11ac, is considered as one of the key enhancements leading WLANs to the Gigabit era. However, as cloud uploading services, Peer-to-Peer (P$2$P) and telepresence applications get popular, the need for a higher uplink capacity becomes inevitable.

In this paper, a unified down/up-link Medium Access Control (MAC) protocol called Uni-MUMAC is proposed to enhance the performance of IEEE 802.11ac WLANs by exploring the multi-user spatial multiplexing technique. Specifically, in the downlink, we implement an IEEE 802.11ac-compliant MU-MIMO transmission scheme to allow the AP to simultaneously send frames to a group of STAs. In the uplink, we extend the traditional one round channel access contention to two rounds, which coordinate multiple STAs to transmit frames to the AP simultaneously. $2$-nd round Contention Window ($CW_{\text{2nd}}$), a parameter that makes the length of the $2$-nd contention round elastic according to the traffic condition, is introduced. Uni-MUMAC is evaluated through simulations in saturated and non-saturated conditions when both downlink and uplink traffic are present in the system. We also propose an analytic saturation model to validate the simulation results. By properly setting $CW_{\text{2nd}}$ and other parameters, Uni-MUMAC is compared to a prominent multi-user transmission scheme in the literature. The results exhibit that Uni-MUMAC not only performs well in the downlink-dominant scenario, but it is also able to balance both the downlink and uplink throughput in the emerging uplink bandwidth-hungry scenario.
\end{abstract}

\section{Introduction}\label{sec:intro}

IEEE 802.11 Wireless Local Area Networks (WLANs) is becoming an indispensable part of our life, at homes and working places. Due to the problems, such as frame collisions and protocol overheads, the throughput of WLANs is significantly lower than the raw data rate of what the Physical (PHY) layer can achieve \cite{ciscoAC}. The evolution of Internet traffic is going to exacerbate this low-throughput problem. The Internet traffic shifts from web browsings and file transfers to a wide variety of applications, many of which integrate content-rich files provided by users \cite{DBLP:conf/icumt/KihlOLA10,DBLP:journals/telsys/WamserPSHT11}. This shift, mainly driven by the bandwidth-hungry multimedia applications (e.g., web HDTV, video sharing and wireless display), demands a performance increase in both downlink and uplink of WLANs \cite{ciscoWhiteP}.

Spatial multiplexing is one of the current trends (the spatial diversity and the frame aggregation are among others) aiming at improving the performance of wireless systems. IEEE 802.11n \cite{5307322} supports spatial multiplexing in the point-to-point communication mode (i.e., Single-user MIMO or SU-MIMO). The point-to-multipoint communication mode, for example, the transmission from the Access Point (AP) to multiple stations (STAs) (i.e., downlink Multi-user MIMO or MU-MIMO), is supported by the latest IEEE amendment-802.11ac \cite{6359961}. However, the uplink MU-MIMO enhancement, which is crucial to mitigate collisions and to satisfy the performance requirements in the uploading-intensive scenario, has not been considered by any IEEE standard. 

In this paper, we propose a unified down/up-link MU-MIMO Medium Access Control (MAC) protocol called Uni-MUMAC, which coordinates distributed STAs to exploit the spatial multiplexing gain to improve the performance of IEEE 802.11ac WLANs. The main contributions are summarized as follows. 1) Two separate MU-MIMO MAC protocols, one for the downlink transmission \cite{5733223} and the other one for the uplink transmission \cite{6314214}, are integrated into a unified MU-MIMO MAC protocol. Compared to \cite{5733223} and \cite{6314214}, where only one-way traffic is considered (i.e., the downlink or the uplink), the presence of both downlink and uplink transmissions has been taken into account. 2) A special focus is placed at finding the most suitable value of the $2$-nd round Contention Window ($CW_{\text{2nd}}$) to obtain the highest system throughput, and the impact of the optimized uplink transmission on the downlink is discussed. With the optimized $CW_{\text{2nd}}$ and other properly configured parameters (e.g., the number of aggregated frames and the queue length of the AP), Uni-MUMAC is then extensively evaluated through simulations in the downlink-dominant and the down/up-link balanced traffic scenarios in IEEE 802.11ac based WLANs. 3) An analytic model is developed to validate the simulation results, and a prominent proposal in the literature is implemented to compare with our scheme. 

The rest of the paper is organized as follows. First, Section \ref{sec:rel_work} explores some of the key MU-MIMO MAC proposals in the literature. Then, Section \ref{sec:mac_protocol} introduces the modified frame structure and detailed Uni-MUMAC operating procedures. After that, Section \ref{sec:perf_eval} gives the considered scenarios to evaluate Uni-MUMAC, the saturation throughput model, simulation results and observations. Finally, Section \ref{sec:conclusions} concludes the paper and discusses the future research challenges.

\section{Related Work}\label{sec:rel_work}

Most previous work has put efforts on adjusting MAC parameters or extending MAC functions to improve the performance of WLANs. In the downlink, the spatial multiplexing technique has recently gained much attention. To support it, many proposals in the literature adopt the following MAC procedure. The AP firstly sends out a modified Request to Send (RTS) containing a group of targeted STAs, then those listed STAs estimate the channel, add the estimated Channel State Information (CSI) into the extended Clear to Send (CTS) and send it back. As soon as the AP receives all successful CTSs, it precodes the outgoing signals and sends multiple data frames simultaneously.

Cai et al. in \cite{conf/globecom/CaiSZSMW08} propose a distributed MU-MIMO MAC protocol that modifies RTS and CTS frames to estimate the channel, based on which, the AP is able to concurrently transmit frames to multiple STAs. Kartsakli et al. in \cite{DBLP:conf/icc/KartsakliZAV09} consider an infrastructured WLAN and propose four multi-user scheduling schemes to simultaneously transmit frames to STAs. The results show that the proposal achieves notable gains compared to that of the single user case. Gong et al. in \cite{DBLP:conf/globecom/GongPSWM10} propose a modified Carrier Sense Multiple Access with Collision Avoidance (CSMA/CA) protocol with three different ACK-replying mechanisms. The authors claim that the proposed protocol can provide a considerable performance improvement against the beamforming based approach when Signal-to-noise Ratio (SNR) is high. Zhu et al. in \cite{DBLP:conf/ccnc/ZhuBKAN12} investigate the required MAC modifications to support downlink MU-MIMO transmissions by focusing on the fairness issue. The proposed Transmit Opportunity (TXOP) sharing scheme not only obtains a higher throughput but is also more fair than the conventional mechanism. Cha et al. in \cite{6214021} compare the performance of a downlink MU-MIMO scheme with a Space Time Block Coding (STBC) based frame aggregation scheme. The results show that the former produces a higher throughput than the latter if transmitted frames are of similar length.

The uplink enhancement is getting more attention as the popularity of P$2$P and cloud applications increases. In general, there are two broad categories of uplink MU-MIMO MAC enhancements, namely, the un-coordinated access and the coordinated access. The former utilizes the MAC random mechanism to decide which STAs are allowed for data transmissions, while the latter employs the AP to schedule STAs' uplink access.

Some of the un-coordinated uplink access schemes are sampled as follows. In \cite{DBLP:conf/wcnc/JinJHS08}, Jin et al. evaluate the performance of uplink MU-MIMO transmissions in the IEEE 802.11 basic access mode, where the simultaneous uplink transmissions are on the random access basis and the channel coefficients of each STA are assumed to be known by the AP. In \cite{4025043}, Zheng et al. present a Distributed Coordination Function (DCF) enhancement called Two-Round RTS Contention (TRRC) to take advantage of the spatial domain. The proposed scheme allows STAs to contend for the channel after a successful RTS is detected. In \cite{kutse}, Tan et al. present a distributed MAC scheme called Carrier Counting Multiple Access (CCMA), where a beacon that contains the uplink access threshold is announced by the AP periodically. Based on the threshold, STAs count the number of ongoing transmissions by monitoring preambles, and then decide to contend for the channel or stay idle. In \cite{BabichC10}, Babich et al. investigate the theoretical model of asynchronous frame transmissions, where a STA is allowed to transmit even if other STAs are already transmitting.

Some of the coordinated uplink access schemes are overviewed as follows. In \cite{ttanc}, Tandai et al. propose a synchronized uplink transmission scheme coordinated by the AP. On receiving requests from STAs, the AP broadcasts a pilot-Requesting CTS (pR-CTS) to schedule STAs' pilot transmissions for estimating the channel. After obtaining the CSI, the AP sends a Notifying-CTS (N-CTS) to inform the selected STAs to transmit frames in parallel. In \cite{SZhou}, Zhou et al. propose a two-round channel contention mechanism, which divides the MAC procedure into two parts, namely, the random access and the data transmission. The random access terminates when the AP receives a predefined number of successful RTSs, and then the data transmission follows. In \cite{journals/twc/Zhang10}, Zhang et al. further extends the two contention rounds to multiple rounds, which enable more STAs to be involved in parallel uplink transmissions. The proposed protocol can fall-back to the single-round mode automatically on condition that the traffic is low and the single-round scheme can provide higher throughput. In \cite{6302115}, Jung et al. present an asynchronous uplink Multi-Packet Reception (MPR) scheme, where an additional feedback channel is assumed to be employed by the AP to acknowledge the successful frame receptions along with other ongoing transmissions.

Only a few work has combined the downlink and the uplink transmissions together. In \cite{DBLP:conf/macom/ShenLSDWZ12}, Shen et al. propose a High Throughput MIMO (HT-MIMO) MAC protocol, which utilizes frequency signatures to differentiate simultaneously-received control messages. The proposal works in the Point Coordination Function (PCF) mode, hence both downlink and uplink transmissions can be only initiated by the AP. In \cite{4917548}, Jin et al. focus on the unbalanced throughput problem between downlink and uplink, where a Contention Window (CW) adjustment scheme and a random piggyback scheme are proposed to increase the downlink throughput ratio. In \cite{DBLP:conf/wcnc/LiAL10}, Li et al. propose a multi-user transmission MAC scheme, which supports the Multi-Packet Transmission (MPT) in the downlink and multiple control frame receptions (e.g., CTSs or ACKs) in the uplink, while simultaneous data transmissions from multiple STAs are not considered. Due to the simplicity, the MAC scheme of \cite{DBLP:conf/wcnc/LiAL10} is implemented to compare with our proposal.

\vspace{-0.9 em}
\section{Uni-MUMAC Operations}\label{sec:mac_protocol}
Uni-MUMAC is based on the IEEE 802.11 Enhanced Distributed Channel Access (EDCA), which relies on the CSMA/CA mechanism to share the wireless channel. EDCA can operate in either the basic access mode or the optional RTS/CTS handshaking one. In this paper, Uni-MUMAC adopts and extends the RTS/CTS scheme for the following reasons: 1) The AP can notify the uplink contending STAs about the number of available antennas by a modified control frame; 2) The AP can estimate the CSI from the RTS/CTS exchanging process; 3) The distributed STAs can be synchronized from the exchanging process to transmit to the AP in parallel.

\subsection{Frame Structure}
\subsubsection{PHY Frame Structure}
\begin{figure}[h!!!!!!!!]
\centering
\includegraphics[scale=0.6]{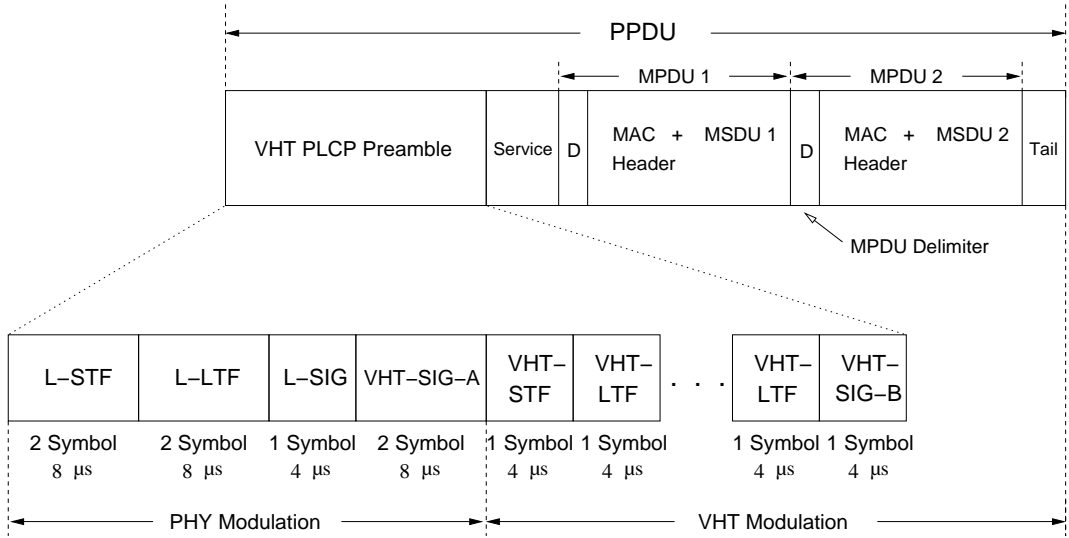}
\caption{PHY frame format of IEEE 802.11ac}
\label{Fig:phy_frame}
\end{figure}

The PHY frame structure of IEEE 802.11ac is shown in Figure \ref{Fig:phy_frame}, where VHT PLCP, PPDU and MPDU stand for Very High Throughput Physical Layer Convergence Protocol, PLCP Protocol Data Unit and MAC Protocol Data Unit, respectively. As shown from the frame structure, PPDU consists of the PHY preamble and MPDUs. IEEE 802.11ac specifies that all MPDUs must be transmitted in the format of Aggregated-MPDU (A-MPDU), where aggregated MPDUs are separated by MPDU delimiters. Before being delivered to the PHY layer, a service field and a tail field are appended to the A-MPDU. The PHY preamble is formed by $3$ legacy fields for the backward compatibility (i.e., L-STF, L-LTF and L-SIG) and some newly introduced VHT fields \cite{6359961}\cite{6140087}.

IEEE 802.11ac introduces these VHT fields to assist WLANs in obtaining the high performance. A Group Identifier (Group-ID) field is added in VHT Signal Field-A (VHT-SIG-A), which is used to inform the targeted STAs about the followed MU-MIMO transmission, the order and the position of each STA's corresponding stream. A complete Group-ID table is created and disseminated by the AP, and will be recomputed as STAs associate or de-associate to the AP. Since the number of STAs' combinations can exceed the available number of Group-ID in a large basic service set, and the down/up-link channel may be different, thus, we assume a single Group-ID can reference to multiple transmission sets along with other PHY preamble features that could be used to resolve the intended STAs \cite{aboul2013managing}. In other words, there will be always at least one proper Group-ID entry that can be mapped to the intended transmission set. 

VHT Long Training Field (VHT-LTF) can contain an orthogonal training sequence that is known by both the transmitter and the receiver to estimate the MIMO channel. The number of VHT-LTF fields should not be less than the number of transmitted spatial streams to precisely estimate the channel. The legacy and VHT-SIG-A fields adopt the low rate modulation scheme to make the preamble understandable to all STAs, while the rest VHT fields and A-MPDU are transmitted using the VHT modulation scheme. In this paper, a single modulation and coding scheme (MCS), i.e., $16$-QAM with $1/2$, is utilized for all frames to simplify the simulation, although the extension to various MCS for different frames and STAs is straightforward. Here, we only introduce the PHY features that are closely related to the proposed protocol. The readers please refer to  \cite{6359961} for details of other PHY features. 
\subsubsection{MAC Frame Structure}
\begin{figure}[h!!!!!!]
\begin{center}
\includegraphics[scale=0.46]{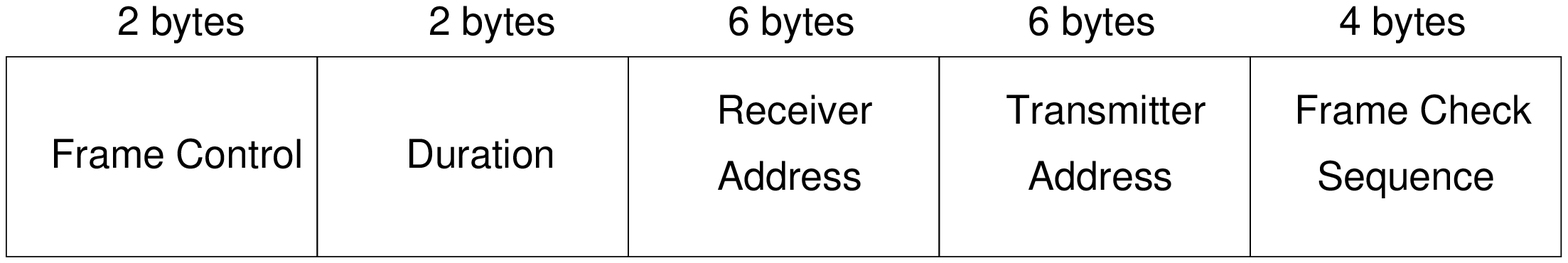}
\caption{Frame structure of standard RTS}\label{Fig:Dframe}
\end{center}
\end{figure}

\begin{figure}[h!!!!!!]
\centering
\subfigure[Ant-CTS]{\includegraphics[scale=0.46]{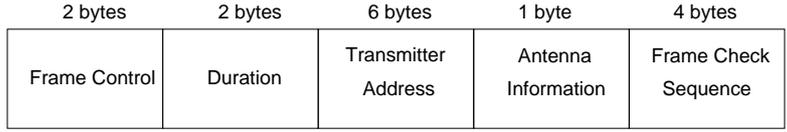}{\label{Fig:uantcts}}}\\ 
\subfigure[G-CTS \& G-ACK]{\includegraphics[scale=0.46]{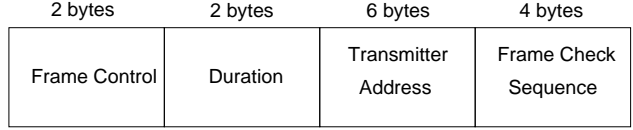}{\label{Fig:uctsack}}} 
\caption{Modified frames for uplink transmissions}\label{Fig:Uframe}
\end{figure}

The control frames of Uni-MUMAC are shown in Figures \ref{Fig:Dframe} and \ref{Fig:Uframe}. In the downlink, the control frames are MU-RTS, MU-CTS and MU-ACK. MU-RTS keeps the standard RTS frame structure, because the AP can utilize the Group-ID field of the PHY frame to notify targeted receivers. MU-CTS and MU-ACK add a transmitter address field to the original CTS and ACK frames in order to facilitate the AP to differentiate multiple responding STAs. Note that MU-CTS and MU-ACK coincidentally have the same frame structure as the standard RTS frame after adding a transmitter address field to the original CTS and ACK frames.

In the uplink, all frame modifications are limited to the AP side to reduce STAs' computing consumption. These modified frames are Ant-CTS (CTS with antenna information), G-CTS (Group CTS) and G-ACK (Group ACK), as shown in Figure \ref{Fig:Uframe}. An antenna information field is added to Ant-CTS, which is broadcast by the AP to announce the number of available antennas (after one antenna is occupied in the first contention round) and the start of the $2$-nd contention round. G-CTS and G-ACK have the identical frame structure, where the receiver address field is removed and replaced by the Group-ID field in the IEEE 802.11ac PHY frame, while a transmitter address field is added to indicate the AP address. The G-CTS frame is used to inform STAs the start of the data transmission, and G-ACK is used to indicate the successful reception of data frames. 
\subsection{Successful Downlink Transmissions}
\begin{figure}[h!!!!!!!!]
\centering
\includegraphics[scale=0.55]{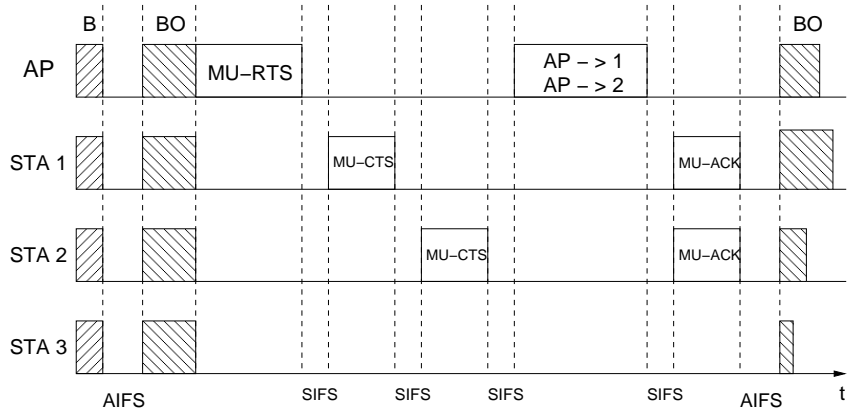}
\caption{A successful Uni-MUMAC downlink transmission}
\label{Fig:smu_down}
\end{figure}

Figure \ref{Fig:smu_down} shows a successful Uni-MUMAC downlink transmission. Initially, the channel is assumed busy (B). After the channel has been idle for an Arbitration Inter Frame Space (AIFS), a random backoff (BO) drawn from CW starts to count down and is frozen as soon as the channel is detected as busy.

Suppose the AP first wins the channel contention and sends a MU-RTS. Then, the STAs who are included in Group-ID reply with MU-CTSs sequentially as the indicated order. Those STAs who are not included in the MU-RTS will set the Network Allocation Vector (NAV) to defer their transmissions. After a MU-CTS is received, the AP will measure the channel through the training sequence included in the PHY preamble, and then uses the estimated CSI to precode the simultaneously-transmitted frames. As being precoded, the frames destined to different STAs will not interfere with each other. Finally, STAs send MU-ACKs at the same time to acknowledge the successful reception of data frames. 

Note that, the uplink channel is assumed to be the same as the downlink one in this paper. In other words, the implicit CSI feedback, namely, the AP estimates the channel using the training sequence included in the MU-CTS, is adopted. The reason is that the explicit CSI feedback will need more computing capability at STAs and require an extra field with substantial volume in the MU-CTS to include the measured CSI, which may not be suitable for STAs in some capacity or power constraint scenarios.

\subsection{Successful Uplink Transmissions}
In the uplink, a standard RTS is sent to the AP by the STA that won the $1$-st round channel contention. Instead of replying a CTS, an Ant-CTS is broadcast by the AP with two functions: 1) to notify the STA about the successful reception of the RTS, and 2) to inform other STAs that the number of available antennas and the start of the $2$-nd contention round. The STAs who have frames to send will compete for the available spatial streams in the $2$-nd contention round. A new random $BO$ ($BO_{\text{2nd}}$) drawn from  $[0,CW_{\text{2nd}}-1]$ starts to count down, and a RTS will be sent if $BO_{\text{2nd}}$ of a STA reaches $0$. The number of available antennas of the AP decreases by one each time an uplink RTS is successfully received. The $2$-nd contention round finishes as: 1) all available antennas of the AP are occupied or 2) a predefined duration of the $2$-nd contention round elapses in case there are not enough contending STAs (the maximum duration of the $2$-nd contention round is set to $CW_{\text{2nd}}$ slots). As soon as the $2$-nd contention round finishes, a G-CTS is sent by the AP to indicate the readiness for receiving multiple frames in parallel. The G-CTS frame includes the STAs who have successfully sent RTSs during both $1$-st and $2$-nd contention rounds. When the G-CTS is received by the targeted STAs, they are synchronized to send data frames to the AP. Finally, the AP acknowledges the received data frames with G-ACK.

\begin{figure}[h!!!!!!!!]
\centering
\includegraphics[scale=0.5]{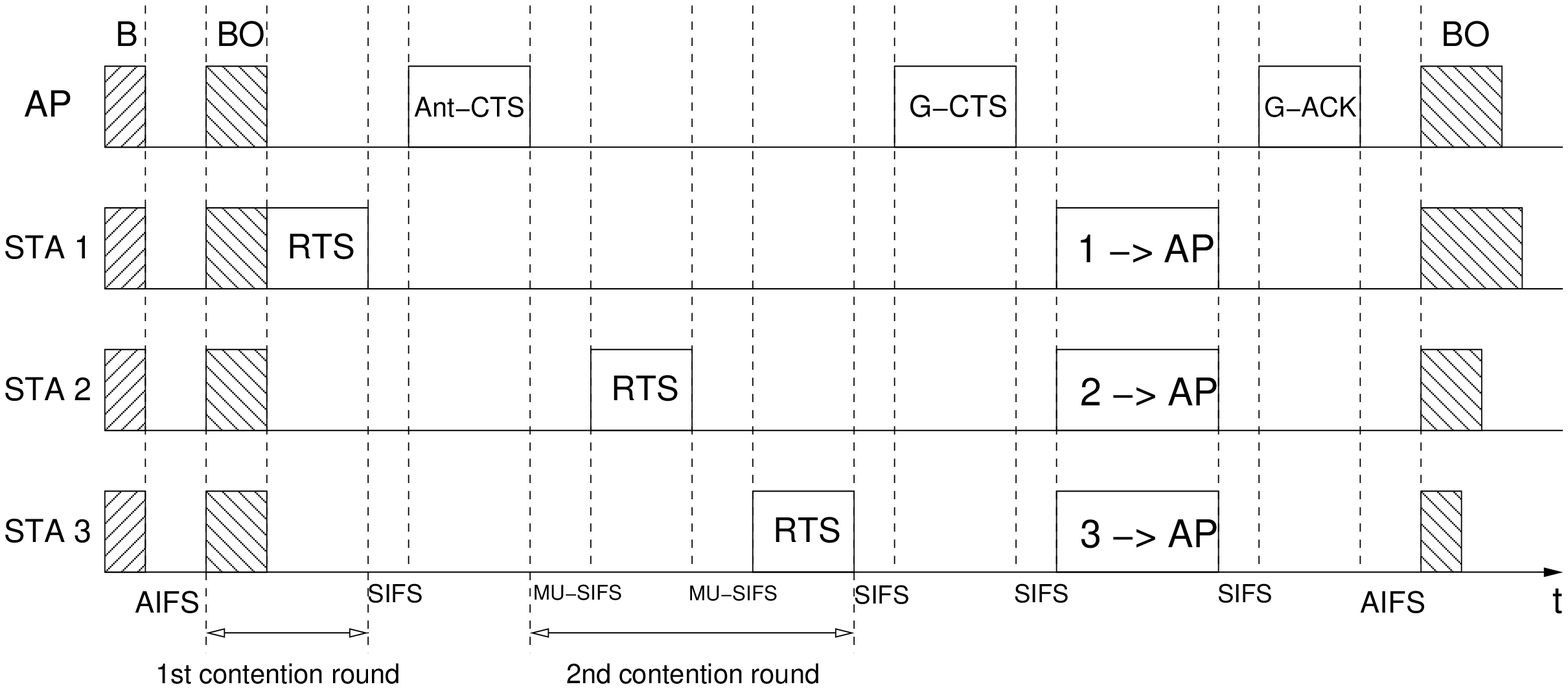}
\caption{A successful Uni-MUMAC uplink transmission} 
\label{Fig:smu_up}
\end{figure}

An example of a successful uplink transmission is shown in Figure \ref{Fig:smu_up}, in which illustrating case, the AP has $3$ antennas, STA $2$ picks $BO_{\text{2nd}}=0$ and STA $3$ picks $BO_{\text{2nd}}=1$ from $[0,CW_{\text{2nd}}-1]$, respectively.

It is important to point out that the RTSs sent by STAs in the $2$-nd contention round could collide with G-CTS sent by the AP. For example, in the case that the RTS sent by a STA who claims the AP's last available antenna is not heard by some STAs (hidden terminals), which therefore believe that the AP still has available antennas. Then, after a Short Inter Frame Space (SIFS) interval, the G-CTS sent by the AP and RTSs sent by the hidden STAs would collide. To avoid this unexpected scenario, STAs are forced to wait for a Multi-User SIFS interval in the $2$-nd contention round. MU-SIFS is an interval longer than SIFS but shorter than AIFS, which not only prioritizes the AP to send the G-CTS, but also avoids STAs to misunderstand MU-SIFS as an idle channel.

\subsection{Frame Collisions}
Collisions will occur in both $1$-st and $2$-nd contention rounds if more than one STA choose the same random backoff value. On sending a RTS, EDCA specifies that the STA has to set a timer according to Equation (\ref{eq:cts}) to receive the expected CTS, where $T_{\text{CTS}}$ represents the transmission duration of a CTS frame. If CTS is not received before the timer expires, the STAs who previously sent RTSs assume that collisions occurred. These RTS-sending STAs will compete for the channel access after the expiration of the timer. For the RTS-receiving STAs, none of RTSs can be decoded correctly. Therefore, after the collision time, the receiving STAs will wait for an Extended Inter Frame Space (EIFS, as shown in Equation (\ref{eq:eifs})) interval to compete for the channel access together with those RTS-sending STAs. 

\begin{figure}[h!!!!!!!!]
\centering
\includegraphics[scale=0.45]{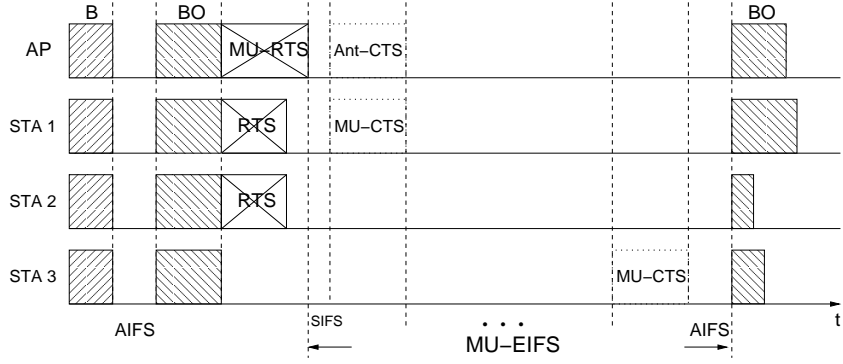}
\caption{Collisions in the $1$-st contention round}\label{Fig:collision}
\end{figure}

As shown in Figure \ref{Fig:collision} (Ant-CTS and MU-CTSs with dotted lines mean these frames would be transmitted if there were no collisions), collisions in the $1$-st contention round include two cases: 1) collisions among STAs; 2) collisions between STAs and the AP. Since STAs can not differentiate these two cases, the collision time has to be set according to the duration of the longer frame, which is $T_{\text{MU-RTS}}$. In addition, the $\text{CTS}_{\text{timer}}$ and the EIFS interval also have to be extended according to $\text{MU-CTS}_{\text{timer}}$ (as shown in Equation (\ref{eq:mu-cts}), where $N$ is the number of AP's antennas) and Multi-User EIFS (MU-EIFS, as shown in Equation (\ref{eq:mu-eifs})), to take the scenario that the AP is involved in collisions into account.

\begin{equation}
\begin{aligned}
  \text{CTS}_{\text{timer}} = \text{SIFS} + T_{\text{CTS}}
\label{eq:cts}
\end{aligned}
\end{equation}
\begin{equation}
\begin{aligned}
  \text{EIFS}= \text{SIFS} + T_{\text{CTS}} + \text{AIFS}
\label{eq:eifs}
\end{aligned}
\end{equation}
\begin{equation}
\begin{aligned}
  \text{MU-CTS}_{\text{timer}} =  N \cdot ( \text{SIFS} + T_{\text{MU-CTS}})
\label{eq:mu-cts}
\end{aligned}
\end{equation}
\begin{equation}
\begin{aligned}
  \text{MU-EIFS} = N \cdot ( \text{SIFS} +  T_{\text{MU-CTS}}) + \text{AIFS}
\label{eq:mu-eifs}
\end{aligned}
\end{equation}

\begin{figure}[h!!!!!!!!]
\centering
\includegraphics[scale=0.5]{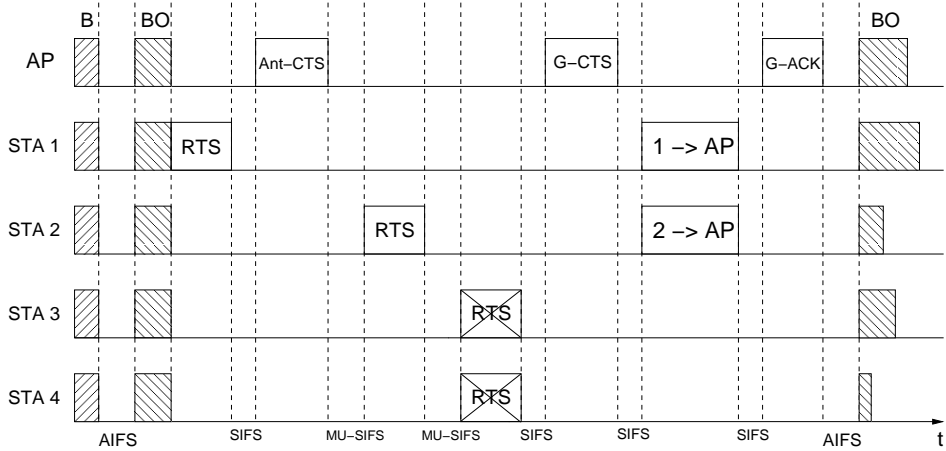}
\caption{RTS collisions in the $2$-nd contention round}\label{Fig:up_collide}
\end{figure}

If collisions occur in the $2$-nd contention round, the colliding STAs will not be indicated as the receivers in the Group-ID field of G-CTS. Therefore, only the STAs that have successfully sent RTSs in both contention rounds are allowed to transmit frames to the AP at the same time (as illustrated in Figure \ref{Fig:up_collide}).

\subsection{Other Considerations}
In IEEE 802.11 EDCA, a STA renews its $BO$ if the channel contention was successful. For the STAs who did not win the contention, the frozen $BO$ is used for the next contention round. In this paper, $BO$ of the $1$-st contention round is renewed after collisions in the $1$-st round or if the STA is the initiator of the two-round process. Although both STA $1$ and STA $2$ participate in the transmission as shown in Figure \ref{Fig:up_collide}, STA $1$ is considered to be the initiator. In other words, STA $1$ will have a new random $BO$ in the followed $1$-st contention round, while STA $2$ will use the frozen $BO$.

It is more straightforward regarding the $BO_{\text{2nd}}$ renewal policy. Each STA draws a fresh $BO_{\text{2nd}}$ from $CW_{\text{2nd}}$ as soon as a new $2$-nd contention round starts.

G-CTS is sent out by the AP when the number of available antennas reaches zero or the duration of the $2$-nd contention round drains. As soon as the Ant-CTS is sent, the AP sets the G-CTS timer to account for up to $CW_{\text{2nd}}$ slots (as shown in Equation (\ref{eq:gctsTimer})).

\begin{equation}
\begin{aligned}
  \text{G-CTS}_{\text{timer}} = CW_{\text{2nd}} \cdot (\text{MU-SIFS} + T_{\text{RTS}})
\label{eq:gctsTimer}
\end{aligned}
\end{equation}

\section{Performance Evaluation}\label{sec:perf_eval}

Uni-MUMAC is evaluated using an analytic model and simulations. The analytic model is adapted from Bianchi's saturation throughput model \cite{bianchi2000performance} to support MU-MIMO transmissions in both downlink and uplink. The simulation is implemented in \verb!C++! using the Component Oriented Simulation Toolkit (COST) library \cite{GilChen} and the SENSE simulator \cite{chen2005sense}.

\begin{figure}[h!!!!!!!!]
\centering
\includegraphics[scale=0.63]{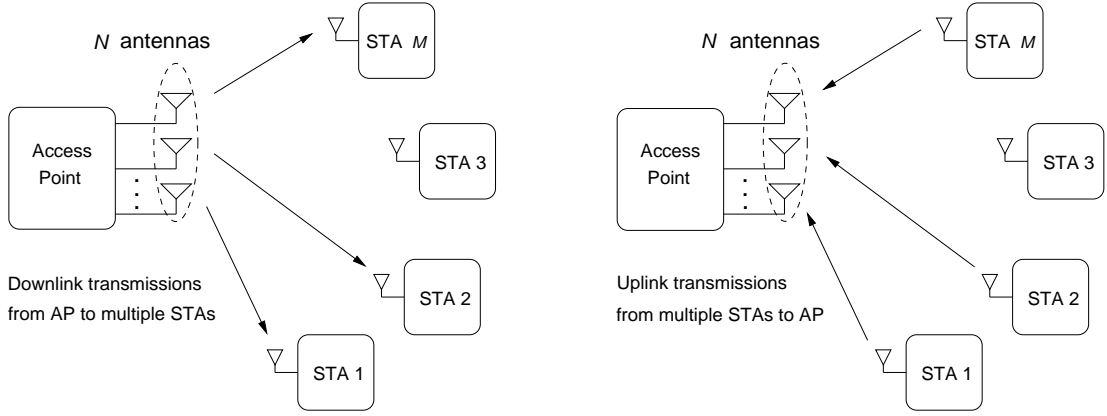}
\caption{Down/Up-link Uni-MUMAC transmissions}\label{Fig:updown_scenario}
\end{figure}

A single-hop WLAN implementing Uni-MUMAC is considered as shown in Figure \ref{Fig:updown_scenario}. It consists of one AP and $M$ STAs with an error-free channel. The AP employs an array of $N$ antennas, while each STA has only one antenna. The data frame has a fixed length of $L$ bits. The parameters used to evaluate Uni-MUMAC are listed in Table \ref{tab_1}.

\subsection{Saturation Throughput Analysis}
\begin{table}[t!!!!!!!]
\centering
\caption{System parameters}
\label{tab_1}
{\normalsize
\begin{tabular}{l l}
\hline\hline
Parameters  &  Values \\ [1ex]
\hline
Channel Bandwidth			& $40$ MHz\\
Modulation \& Coding Scheme			& $16$-QAM with $1/2$\\
Guard Interval 		&	$0.8$ $\mu$s\\
Queue Length of STA \& AP  		&	$Q_{\text{sta}}=50$, $Q_{\text{ap}}=M^2$\\	
Frame Length ($L$)     	&	$8000$ bits\\
MAC Header ($L_{\text{MAC}}$)   		& 	$272$ bits\\
MPDU Delimiter ($L_{\text{delimiter}}$)     & 	$32$ bits\\
Service Bits ($L_{\text{service}}$)  	& 	$16$ bits\\
Tail Bits ($L_{\text{tail}}$)  	& 	$6$ bits\\
RTS/MU-RTS/MU-CTS/MU-ACK 	&   $160$ bits\\
Ant-CTS   	& 	$120$ bits\\
G-CTS/G-ACK 	&   $112$ bits\\
Idle Slot ($\sigma$) 			&  	$9$ $\mu$s\\
SIFS, MU-SIFS and AIFS 				&	$16$, $20$ and $34$ $\mu$s\\
$CW$	  		& 	$32$ \\
AP Antennas ($N$)     	&	$1$, $2$, $4$\\
No. of Iteration ($N_\text{iteration}$) &	$100000$\\
\hline
\end{tabular}
}
\label{table1}
\end{table}

Let $\tau = \frac{2}{(CW+1)}$ be the transmission probability of a node in a random slot, where $CW$ is the size of the $1$-st round contention window. Then, the probability that the channel is idle is:
\begin{equation}
  p_\text{i}=(1-\tau)^{M+1}.
\label{eq:pempty}
\end{equation}

The probability that the channel sees a successful transmission slot, $p_\text{s}$, is given by: 
\begin{align}
  p_\text{s} =\binom{M+1}{1}\tau(1-\tau)^{M}   = (M+1) \tau (1-\tau)^{M},
\label{eq:success1}
\end{align}
which accounts for that a single node (either the AP or a STA) successfully wins the $1$-st round channel contention.

By deducting $p_\text{i}$ and $p_\text{s}$, the probability that the channel observes a collision slot, $p_\text{c}$, is obtained:
\begin{equation}
\begin{aligned}
  p_\text{c}=1-p_\text{i}-p_\text{s}.
    \end{aligned}
\label{eq:pcc}
\end{equation}

In the saturated condition, a successful downlink transmission always contains $N$ (the number of AP antennas) data streams. Therefore, the number of bits of a successful downlink transmission ($N_\text{b,down}$) is:  
\begin{equation}
\begin{aligned}
  N_\text{b,down}= \alpha \cdot N \cdot N_{\text{f}} \cdot L \cdot p_\text{s},
\label{eq:bits-dw}
  \end{aligned}
\end{equation}
where $\alpha=\frac{1}{M+1}$ is the probability that a transmission is from the AP, and $N_{\text{f}}$ is the number of aggregated frames in an A-MPDU.

The calculation of the successfully received number of bits of uplink ($N_\text{b,up}$) has to account for successful transmissions of both $1$-st and $2$-nd contention rounds:
\begin{equation}
\begin{aligned}
  N_\text{b,up}= (1-\alpha)  \cdot N_{\text{f}} \cdot L \cdot p_\text{s} \cdot \sum_{\text{x}=1}^{N} p_\text{x\_ant} \cdot \text{x},
\label{eq:bits-up}
  \end{aligned}
\end{equation}
where $p_\text{x\_ant}$ is the probability that x (x $\in [1, N]$) antennas of the AP have been used for the uplink transmission. In other words, one antenna has been obtained by a STA in the $1$-st contention round, and x-1 antennas have been successfully obtained by STAs in the $2$-nd contention round.

The duration of a successful downlink transmission, $T_{\text{s,down}}$, is:
\begin{equation}
\begin{aligned}
T_{\text{s,down}} = \text{AIFS}
+T_{\text{MU-RTS}} + N\cdot(T_{\text{MU-CTS}} + \text{SIFS}) + T_{\text{A-MPDU}} + T_{\text{MU-ACK}} + 2 \cdot \text{SIFS}.
\label{eq:tsdw}
\end{aligned}
\end{equation}

An example to calculate the duration of a MU-RTS frame and a data frame using the system parameters of Table \ref{tab_1} is given in Equation (\ref{eq:frame_duration}). $T_{\text{PHY}}(N)=36+N\cdot 4$ $\mu$s are the duration of PHY header (the number of the VHT-LTF fields is proportional to the number of AP antennas $N$); $L_{\text{service}}$, $L_{\text{tail}}$ and $L_{\text{delimiter}}$ are the length of the service field, the tail field and the MPDU delimiter; $L_{\text{DBPS}}$ and $T_{\text{symbol}}$ are the number of data bits in a symbol and the symbol duration; $N_{\text{f}}$ is the number of aggregated frames in an A-MPDU; $L_{\text{MU-RTS}}$ and $L_{\text{MAC}}$ are the length of MU-RTS and the MAC header respectively. More detailed calculation of the frame duration can be found in \cite{6287486}.
\begin{equation}
\begin{aligned}
\begin{cases}
  T_{\text{MU-RTS}}=T_{\text{PHY}}(N)+\Big\lceil\frac{L_{\text{service}}+L_{\text{MU-RTS}}+L_{\text{tail}}}{L_{\text{DBPS}}}\Big\rceil T_{\text{symbol}} \\

  T_{\text{A-MPDU}}=T_{\text{PHY}}(N)+\Big\lceil\frac{L_{\text{service}}+ N_{\text{f}}\cdot (L_{\text{MAC}}+ L + L_{\text{delimiter}}) + L_{\text{tail}}}{L_{\text{DBPS}}}\Big\rceil T_{\text{symbol}}
\label{eq:frame_duration}
\end{cases}
\end{aligned}
\end{equation}

The duration of a successful uplink transmission, $T_{\text{s,up}}$, is:
\begin{equation}
\begin{aligned}
T_{\text{s,up}} = \text{AIFS}
+T_{\text{RTS}} + T_{\text{Ant-CTS}} + E_{\text{2nd-slots}}+ T_{\text{G-CTS}} + T_{\text{A-MPDU}} + T_{\text{G-ACK}} + 4 \cdot \text{SIFS},
\label{eq:tsup}
\end{aligned}
\end{equation}
where $E_{\text{2nd-slots}}$ stands for the average duration of the $2$-nd contention round.

\begin{equation}
\begin{aligned}
E_{\text{2nd-slots}}=(T_{\text{RTS}} + \text{MU-SIFS})\cdot \sum_{\text{k}=1}^{CW_\text{2nd}} p_\text{k\_Slot} \cdot \text{k},
\label{eq:tav}
\end{aligned}
\end{equation}
where $p_\text{k\_Slot}$ is the probability that there are k (k $\in [1, CW_\text{2nd}]$) slots in the $2$-nd contention round.

As a STA can not differentiate if collisions of the $1$-st round are caused by the AP or other STAs, the collision time has to be set according to the duration of the longer frame:
\begin{equation}
\begin{aligned}
T_{\text{c}} = \text{AIFS}
+T_{\text{MU-RTS}} + N\cdot(T_{\text{MU-CTS}} + \text{SIFS}).
\label{eq:tc}
\end{aligned}
\end{equation}

The average duration of a channel slot is:
\begin{equation}
\begin{aligned}
T_{\text{average}} = \alpha \cdot p_\text{s} \cdot T_{\text{s,down}}  + (1-\alpha)\cdot p_\text{s} \cdot T_{\text{s,up}}
+p_{\text{c}}\cdot T_{\text{c}} + p_{\text{i}}\cdot \sigma.
\label{eq:te}
\end{aligned}
\end{equation}

Equation (\ref{eq:p2ant}) gives a simple example to calculate $p_\text{2\_ant}$, in which case, the AP has $2$ antennas and $CW_\text{2nd} =2$: 
\begin{equation}
\begin{aligned}
  p_\text{2\_ant}= \binom{M-1}{1}\frac{1}{CW_\text{2nd}}\left(1-\frac{1}{CW_\text{2nd}}\right)^{M-2} + \binom{M-1}{1}\frac{1}{CW_\text{2nd}}\left(1-\frac{1}{CW_\text{2nd}}\right)^{M-2}\cdot p_\text{1\_fail}.
\label{eq:p2ant}
\end{aligned}
\end{equation}

The first part of Equation (\ref{eq:p2ant}) stands for that only one STA is successful in the $1$-st slot. The second part represents that only one STA is successful in the $2$-nd slot, which is conditioned on that the $1$-st slot fails ($p_\text{1\_fail}$, no STAs or more than one STA chooses the $1$-st slot). Note that the similar condition is not required for the first part, because the $2$-nd round contention finishes as soon as a STA wins the $1$-st slot regardless the choices of other STAs of other slots. As $CW_\text{2nd}$ increases, the closed form of $p_\text{2\_ant}$ becomes infeasible due to various combination of conditions for a STA to succeed in different slots. Therefore, we utilize a semi-analytic algorithm to calculate $p_\text{x\_ant}$ and $p_\text{k\_Slot}$, the pseudo code of which is shown in Algorithm \ref{Alg:pant}.


\begin{algorithm}[h!!!!!!!]
\footnotesize
\caption{Algorithm for $p_\text{x\_ant}$ and $p_\text{k\_Slot}$}\label{Alg:pant}
\KwIn{$N$, $M$, $CW_\text{2nd}$, $N_\text{iteration}$}

\KwOut{$p_\text{x\_ant}$, $p_\text{k\_Slot}$}
\BlankLine
\For{$i \leq N_\text{iteration}$}
 {\BlankLine
 		\For{$\textbf{STA\_id} = 1:M$}
 		{\BlankLine
 		$rand\_value(STA\_id)=\lfloor rand()*CW_\text{2nd} \rfloor$\Comment{\textbf{$rand\_value$: random value chosen by each STA}}
 		\BlankLine
		$CW2nd\_Array(rand\_value(STA\_id) )\texttt{++}$\Comment{\textbf{$CW2nd\_Array$: no. of times a $CW_\text{2nd}$ value is chosen}}

 		}
 		\BlankLine
 		$count=1$
 		
 		$1st\_success=\textbf{false}$ 
 		\BlankLine
 				
 		\For{$\textbf{length} = 1:CW_\text{2nd}$}
  		{\BlankLine
  		
  		 \If{$CW2nd\_Array(length) \texttt{==} 1$}
  		    {  		\BlankLine      		    
			 $num\_sta\_success(count)\texttt{++}$\Comment{\textbf{$num\_sta\_success$: no. of STAs won the $2$-nd round contention}}
			 \BlankLine
			 \If{$1st\_success \texttt{==} \textbf{true}$}
  		        {\BlankLine$num\_sta\_success(count\texttt{-}1)\texttt{--}$}

  		     \If{$count \texttt{==} 1$}
  		        {\BlankLine
  		        $1st\_success=\textbf{true}$}
  		        
  		     \If{$count \texttt{==} N\texttt{-}1$}
  		        {\BlankLine\textbf{break}}
\BlankLine
			 	 $count\texttt{++}$

  		
				    	
				    	
  		      
  		    }
  		    \If{$length \texttt{==} CW_\text{2nd}$}
  		      {				    	
				    	\BlankLine
				    	 $slot\_id=CW_\text{2nd}$	
				    	
  		      }
 		}
 		\BlankLine
 		$num_\text{slot\_success}(slot\_id)\texttt{++}$\Comment{$num_\text{slot\_success}$\textbf{: no. of slots used in the $2$-nd round contention}}

}
 		\For{$\text{x} = 1:N\texttt{-}1$}
  		{\BlankLine
  		    $p_\text{(x+1)\_ant}=num\_sta\_success(\text{x})/N_\text{iteration}$	
  		}
  		
  		 $p_\text{1\_ant}=1-\sum_{x=2}^{N} p_\text{x\_ant}$
  		 
\BlankLine
 		\For{$k = 1:CW_\text{2nd}$}
  		{\BlankLine
  		    $p_\text{k\_Slot}=num_\text{slot\_success}(\text{k})/N_\text{iteration}$
  		}
  		
\end{algorithm}

Finally, the collision probability of a node,
\begin{equation}
\begin{aligned}
P_{\text{collision}} = 1-(1-\tau)^M,
\label{eq:pcsta}
\end{aligned}
\end{equation}
and down/up-link throughput are derived:
\begin{equation}
\large
\begin{aligned}
\begin{cases}
S_{\text{down}} = \frac{N_\text{b,down}}{T_{\text{average}} } \\

S_{\text{up}} = \frac{N_\text{b,up}}{T_{\text{average}} }.
\label{eq:sdwup}
\end{cases}
\end{aligned}
\end{equation}

The transmission probability $\tau$, equations (\ref{eq:pcsta}) and (\ref{eq:sdwup}) form a non-linear system, which can be resolved by an iterative numerical technique \cite{kumar2005new}.

\subsection{System Performance against $CW_{\text{2nd}}$}
\begin{figure*}[t!!!!!!]
\begin{center}
\subfigure[$M=8$]{\includegraphics[scale=0.52]{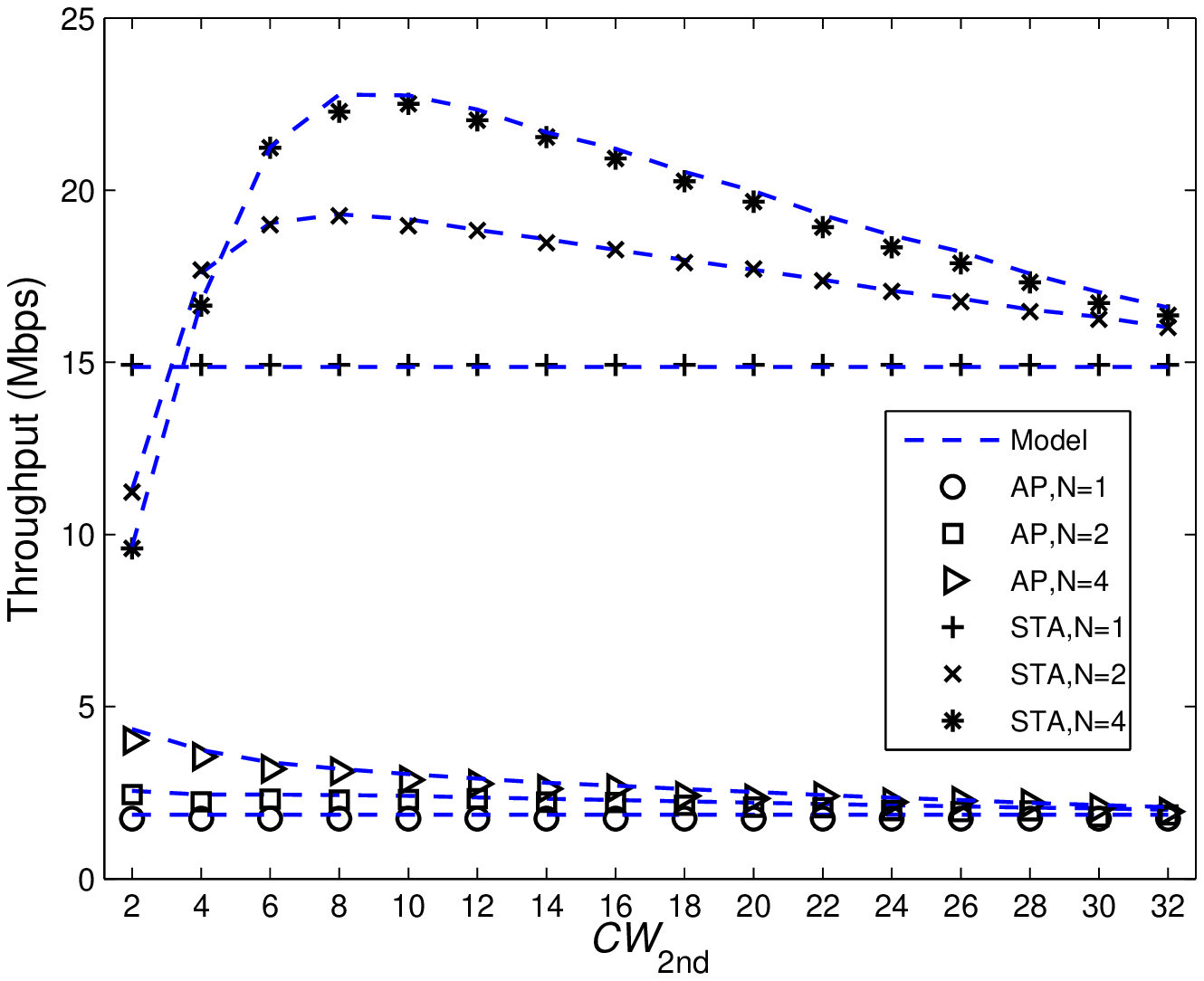}{\label{Fig:tpS_var}}}\hspace{-2 em}
\subfigure[$M=15$]{\includegraphics[scale=0.52]{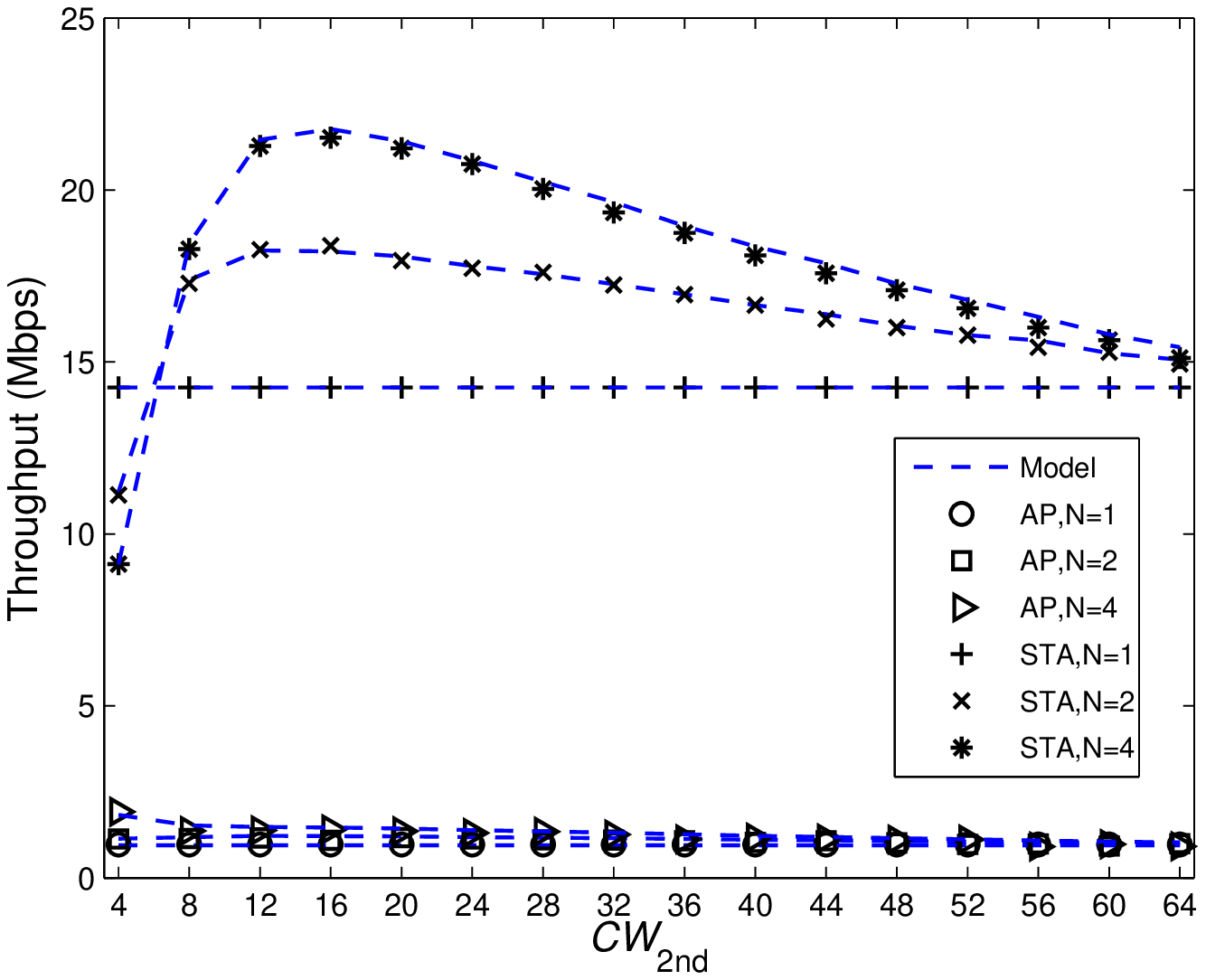}{\label{Fig:tpB_var}}}\\
\caption{Saturated throughput against $CW_{\text{2nd}}$}\label{Fig:tp_var}
\end{center}
\end{figure*}

\begin{figure*}[t!!!!!!]
\begin{center}
\subfigure[$M=8$, STA $1.4$ Mbps, AP $11.2$ Mbps]{\includegraphics[scale=0.52]{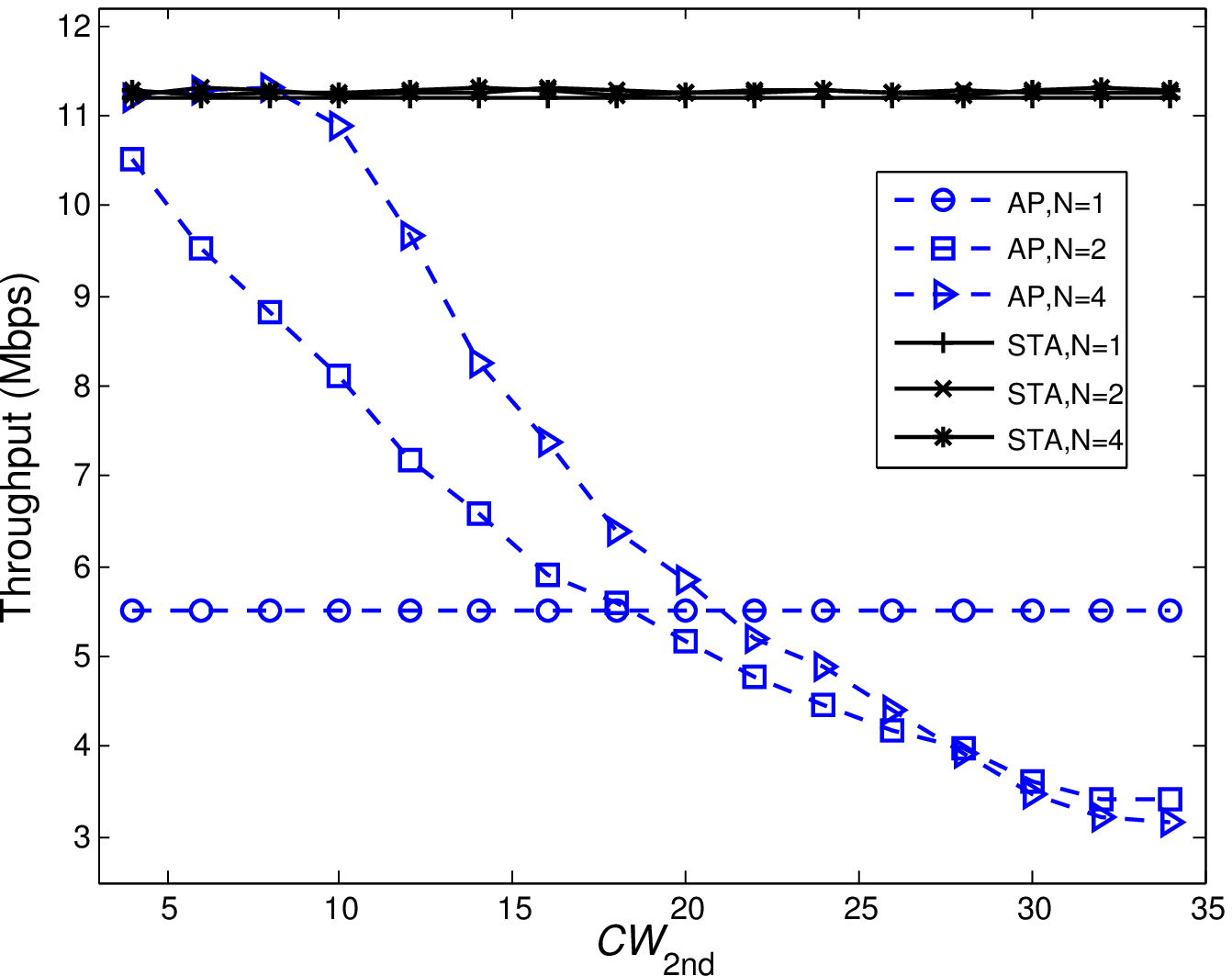}{\label{Fig:tpS_var1}}}\hspace{0.9cm}
\subfigure[$M=8$, STA $1.4$ Mbps, AP $11.2$ Mbps]{\includegraphics[scale=0.52]{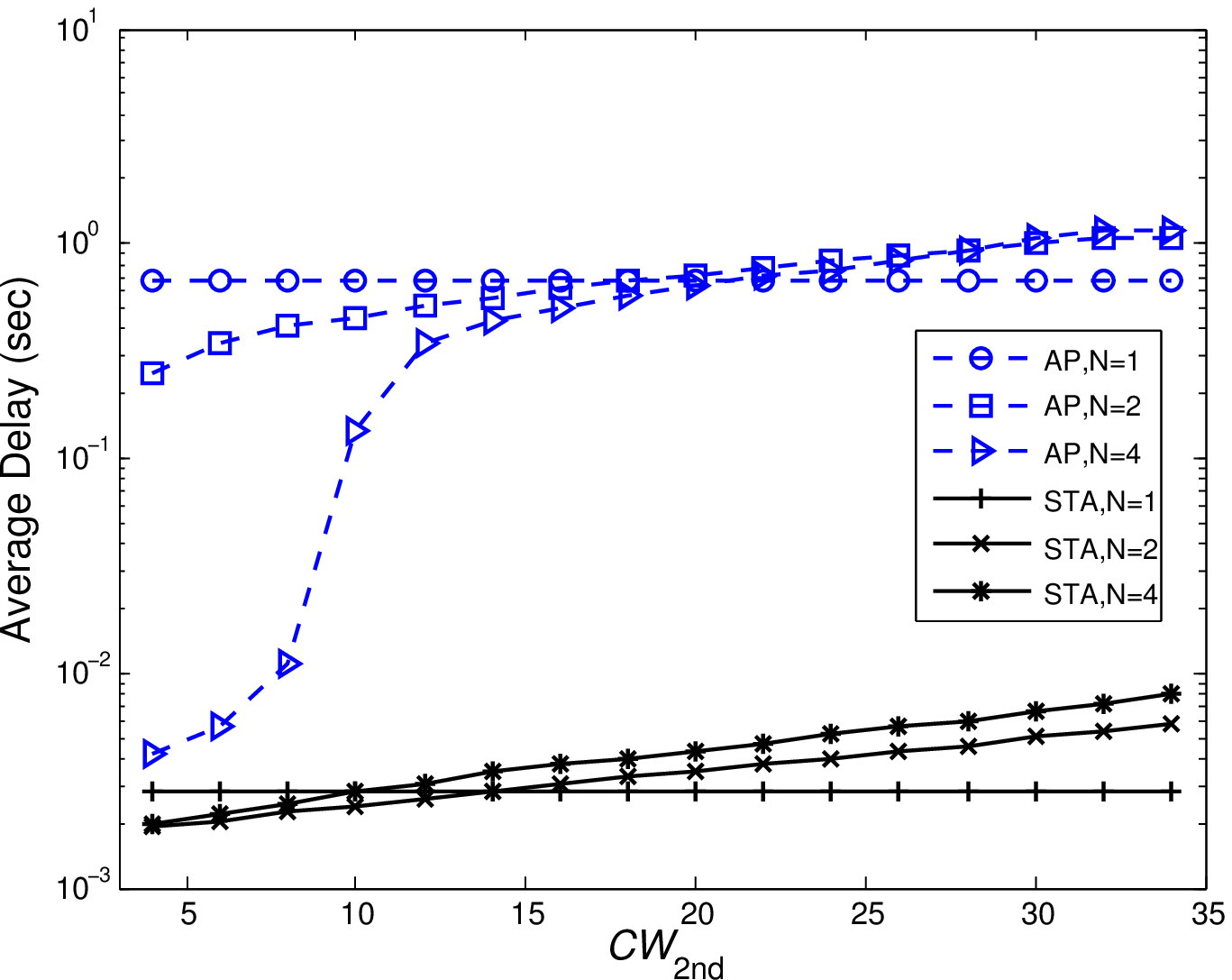}{\label{Fig:tpB_var1}}}\\
\caption{Non-saturated throughput \& Average delay}\label{Fig:tp_var1}
\end{center}
\end{figure*}

In this sub-section, the performance of Uni-MUMAC is evaluated by increasing $CW_{\text{2nd}}$, with the goal to find a suitable $CW_{\text{2nd}}$ value that maximizes the system performance. Two traffic conditions are considered: 1) the saturated one, as shown in Figure \ref{Fig:tp_var}, and 2) the non-saturated one, as shown in Figure \ref{Fig:tp_var1}. The saturated condition means that both the AP and STAs always have frames to transmit. Obviously, there is no $2$-nd round channel access when the AP has $1$ antenna, which is why the results keep constant as $N=1$. Note that the plots include both analysis and simulation results in the saturated condition, while the plots include only simulation results of the non-saturated condition.

As shown in Figure \ref{Fig:tp_var}, when the WLAN is saturated (i.e., both downlink and uplink are saturated), $CW_{\text{2nd}}$ has very small impact on the downlink throughput (AP's throughput). However, for the uplink, the importance of choosing an appropriate $CW_{\text{2nd}}$ is observed. For example, the uplink throughput (STAs' throughput) approaches its maximum when $CW_{\text{2nd}} \in [8,12]$ as $M=8$ (Figure \ref{Fig:tpS_var}) and when $CW_{\text{2nd}} \in [12,16] $ as $M=15$ (Figure \ref{Fig:tpB_var}). 

In the non-saturated condition, we set the traffic load for each STA and the AP to $1.4$ Mbps and $11.2$ Mbps, respectively. In Figure \ref{Fig:tpS_var1}, the downlink throughput ($N=2$ and $4$) obtains the highest value when $CW_{\text{2nd}} \in [4,8]$, and then decreases as $CW_{\text{2nd}}$ keeps increasing. The reason is that the continuous increase of $CW_{\text{2nd}}$ leads to longer uplink transmissions that harm the downlink ones. Figure \ref{Fig:tpB_var1} shows that the average delay increases as $CW_{\text{2nd}}$ increases. Note that, the average delay remains at a relatively low level when the system is in the non-saturated condition, for example, the average delay of STAs when $CW_{\text{2nd}} \in [4,34]$ and the average delay of the AP when $N=4$ and $CW_{\text{2nd}} \in [4,8]$. However, the average delay of the AP ($N=4$) increases sharply as the downlink traffic approaches saturation. 

It is also observed that the downlink throughput, as the network becomes saturated, is much lower than the uplink one. The reasons are as follows. First, the AP bottle-neck effect. It is due to the fact that the AP manages all traffic to and from STAs in a WLAN, while it has the same probability to access the channel as the STAs due to the random backoff mechanism of CSMA/CA. In addition, the inherently high traffic load at the AP results in that the downlink is saturated in most of the time. Thirdly, a favorable value of $CW_{\text{2nd}}$ for the uplink does not mean the same benefit to the downlink. For example, as shown in the Figure \ref{Fig:tp_var}, the uplink obtains the highest throughput when $CW_{\text{2nd}}$ is set approximately to $M$ ($CW_{\text{2nd}} \approx M$), while the downlink transmission prefers a value of $CW_{\text{2nd}}$ as small as possible. 

In order to mitigate the AP bottle-neck effect and compensate the downlink disadvantage when STAs choose a big $CW_{\text{2nd}}$, we set the maximum number of frames that the AP can aggregate in an A-MPDU to $M$ ($N_{\text{f}} \leq M$), while keeping the number of frames aggregated by each STA to $1$ in the following simulations. Also, the queue length of the AP is set to quadratically increase with the number of STAs ($Q_{\text{ap}}=M^2$) to statistically guarantee that there are enough frames destined to different STAs \cite{6287486}.

\begin{figure*}[h!!!!!!]
\begin{center}
\subfigure[$M=8$]{\includegraphics[scale=0.520]{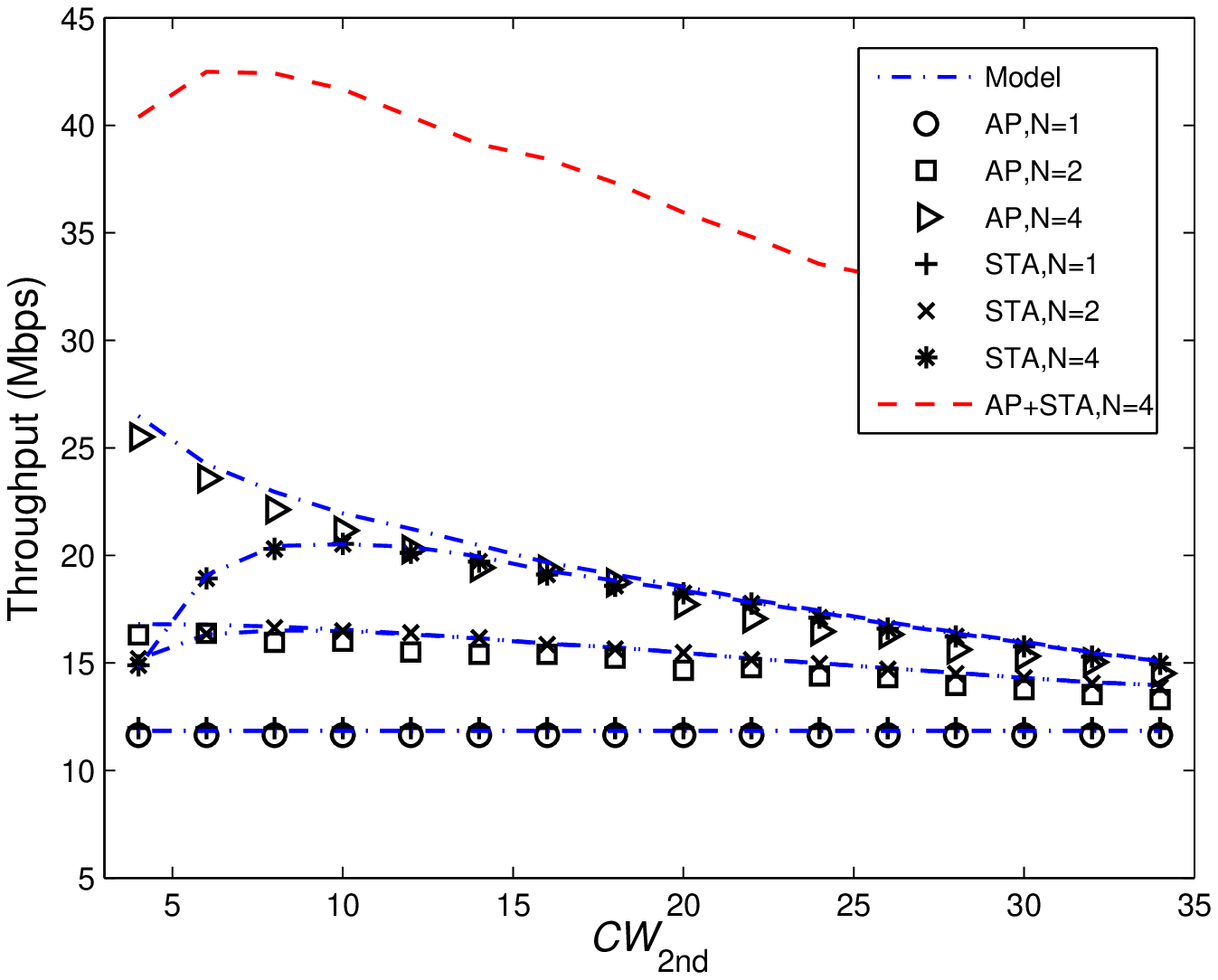}{\label{Fig:tpSA_var}}}\hspace{-2 em}
\subfigure[$M=15$]{\includegraphics[scale=0.520]{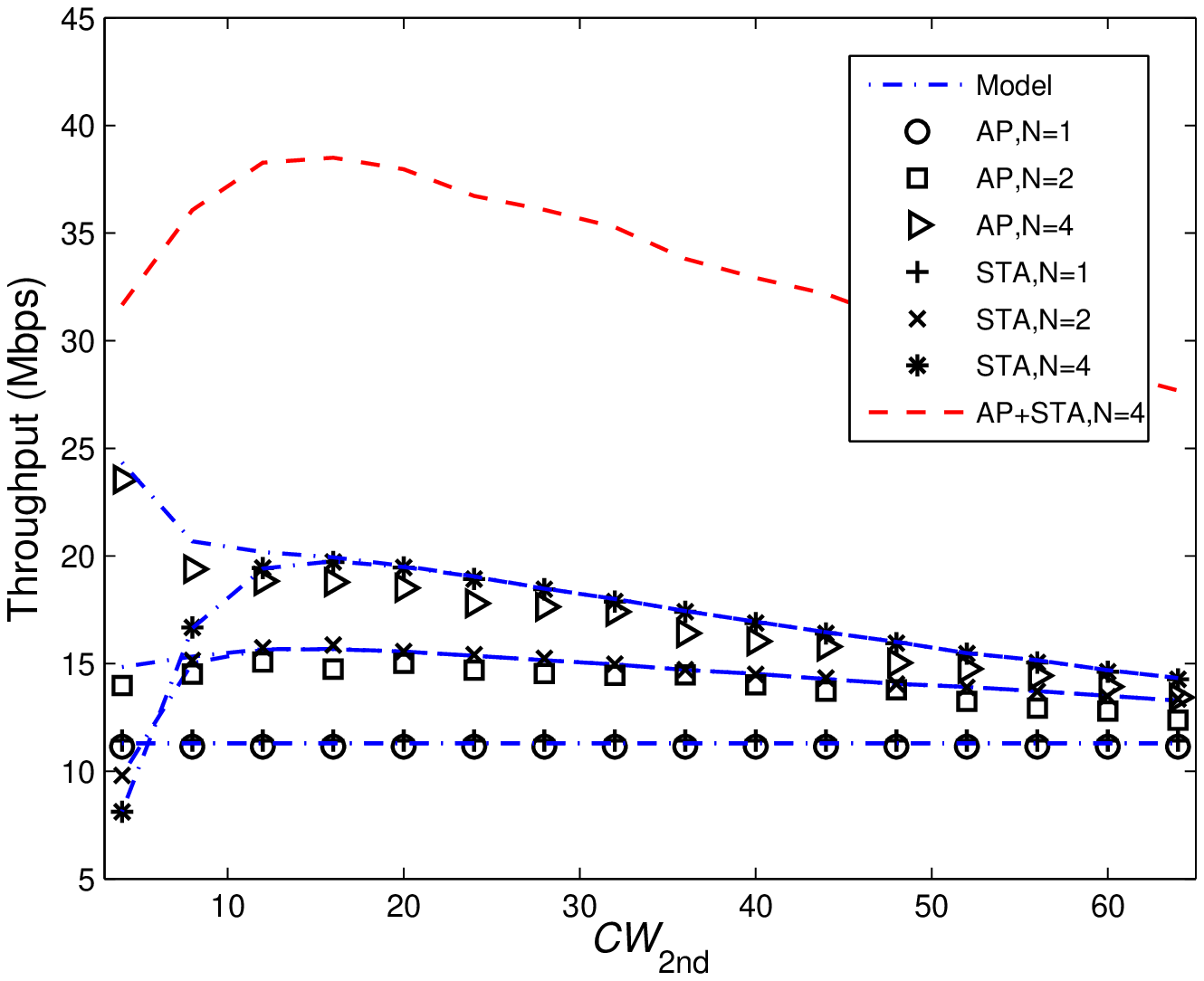}{\label{Fig:tpBA_var}}}\\
\caption{Saturated throughput when AP aggregates frames}\label{Fig:tpA_var}
\end{center}
\end{figure*}

\begin{figure*}[h!!!!!!]
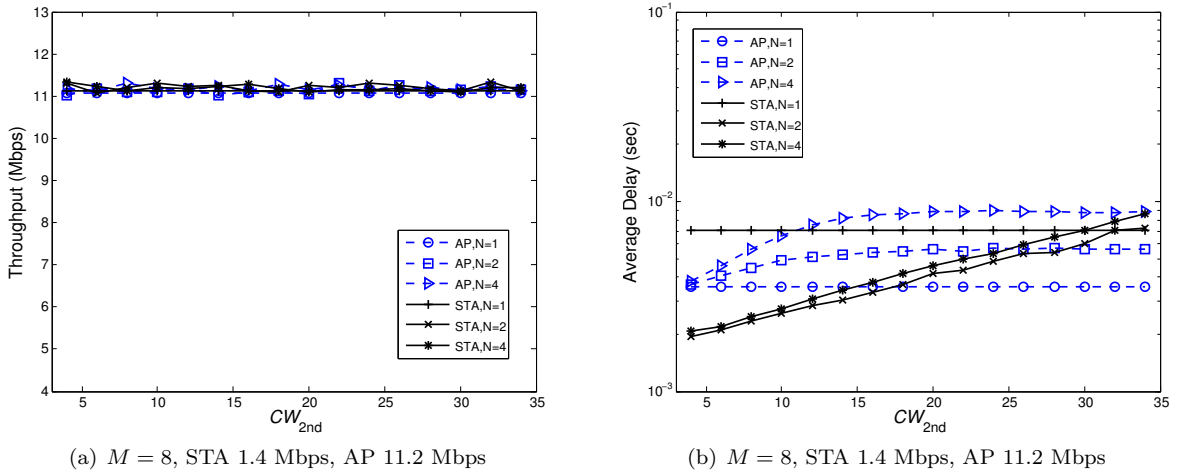

\begin{center}
\subfigure[$M=8$, STA $1.4$ Mbps, AP $11.2$ Mbps]{\includegraphics[scale=0.52]{xfigs/
up_down_sim_tp_1lambda175_n9_CW32_cw4_34_seeds310_apNodM_mn1M_a1_4_NewPHY_03Jul13-crop.eps}{\label{Fig:tpSA_var1}}}\hspace{0.9cm}
\subfigure[$M=8$, STA $1.4$ Mbps, AP $11.2$ Mbps]{\includegraphics[scale=0.52]
{xfigs/
up_down_sim_AVdelay_1lambda175_n9_CW32_cw4_34_seeds310_apNodM_mn1M_a1_4_NewPHY_03Jul13-crop.eps}{\label{Fig:tpBA_var1}}}\\
\caption{Non-saturated throughput \& Average delay when AP aggregates frames}\label{Fig:tpA_var1}
\end{center}
\end{figure*}

In Figures \ref{Fig:tpA_var} and \ref{Fig:tpA_var1}, the performance of Uni-MUMAC is evaluated in the same condition as done in Figures \ref{Fig:tp_var} and \ref{Fig:tp_var1} except that the network adopts the new frame aggregation scheme (AP's $N_{\text{f}} \leq M$, STA's $N_{\text{f}} =1$) and the new queue length ($Q_{\text{ap}}=M^2$, $Q_{\text{sta}}=50$). The results show that Uni-MUMAC manages to avoid the extremely low downlink throughput when the system is saturated (Figure \ref{Fig:tpA_var}) and keeps the downlink transmission always in the non-saturation area (Figure \ref{Fig:tpSA_var1}), which is not achieved in Figure \ref{Fig:tpS_var1}. The average delay of the AP (Figure \ref{Fig:tpBA_var1}) is much lower compared to that of the AP in Figure \ref{Fig:tpB_var1}, which is because the system remains in the non-saturated condition by employing the frame aggregation scheme.

The results from Figure \ref{Fig:tpA_var} also show that the system can roughly obtain the maximum performance when $CW_{\text{2nd}} \in [M-4,M+4]$. For example, in the case that the AP has $4$ antennas, the system throughput (AP+STA) reaches its maximum when $CW_{\text{2nd}} \in [6,8]$ as $M=8$ and $CW_{\text{2nd}} \in [12,16]$ as $M=15$, respectively. Therefore, the optimum value of $CW_{\text{2nd}}$ is fixed to $M$ in the following simulations.

\subsection{System Performance against $M$}

In this sub-section, the performance of Uni-MUMAC is evaluated against the number of STAs in the downlink-dominant and the down/up-link balanced traffic scenarios, where $M$ is increased from $1$ to $35$, the maximum number of frames aggregated at the AP is set to $M$ and the $2$-nd round Contention Window is also set to $M$. The two traffic scenarios are specified as follows.
\begin{enumerate}
\item{\textbf{Downlink-dominant}}: This is the traditional WLAN traffic scenario, where the AP manages a much heavier traffic load compared to that of STAs. Therefore, the traffic load of the AP is set to be $4$ times higher than that of each STA. For instance, if the traffic load of a STA is $0.8$ Mbps and there are $5$ STAs, the traffic load of the AP will be $4 \cdot 0.8 \cdot 5=16$ Mbps.
\item{\textbf{Down/up-link balanced}}: This is one of WLAN traffic types that not only includes P$2$P applications, which have already been around for some years, but also includes those emerging content-rich file sharing and video calling applications. Therefore, the traffic load of the AP is set to be the same as that of each STA. In this case, if there are $5$ STAs, and each STA has $0.8$ Mbps traffic load, the traffic load of the AP will be $0.8 \cdot 5=4$ Mbps.
\end{enumerate}

The multi-user MAC scheme (LI-MAC) proposed by Li et al. in \cite{DBLP:conf/wcnc/LiAL10} is implemented and used as a reference (named as AP/STA-LI in the legend) to compare with Uni-MUMAC. For fair comparison, LI-MAC and Uni-MUMAC adopt the same configuration parameters (as shown in Table \ref{tab_1}). The key features of LI-MAC and Uni-MUMAC are illustrated in Table \ref{tab_2}.
\newcommand{\tabincell}[2]{\begin{tabular}{@{}#1@{}}#2\end{tabular}}
\begin{table}[h!!!!!!!]
\newsavebox{\tablebox}
\caption{{Key features of LI-MAC and Uni-MUMAC}}
\label{tab_2}
\begin{lrbox}{\tablebox}
\renewcommand\arraystretch{2.5}
{\LARGE
\begin{tabular}{|c|c|c|c|c|}
\hline
{\bfseries \Large MAC Schemes}  & {\bfseries \Large Modification}  & {\bfseries \Large Downlink} & {\bfseries \Large Uplink} \\ \hline 
LI-MAC \cite{DBLP:conf/wcnc/LiAL10} & MU-RTS/CTS & Multi-packet + Parallel-control-frame TX & One-round single-packet TX  \\ \hline %

Uni-MUMAC & \tabincell{l}{\vspace{-0.2em}MU-RTS/CTS/ACK/SIFS\\ \vspace{-0.01em}G-CTS/ACK, Ant-CTS}& Multi-packet + Sequential-control-frame TX & Two-round Multi-packet TX  \\ \hline %

\end{tabular}}
\label{upmac}
\end{lrbox}
\resizebox{0.99\textwidth}{!}{\usebox{\tablebox}}
\end{table}

\begin{figure*}[t!!!!!!]
\begin{center}
\subfigure[Downlink-dominant: STA $0.8$ Mbps, AP $3.2$ Mbps]{\includegraphics[scale=0.52]{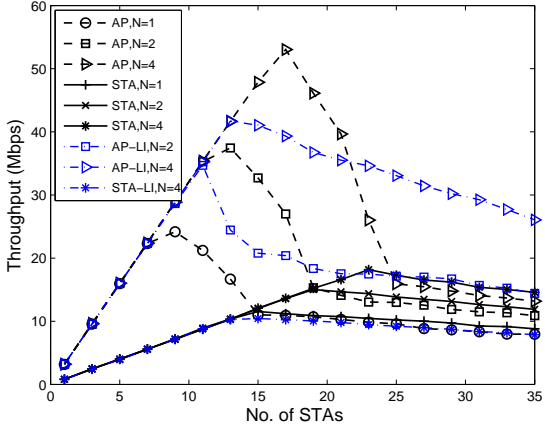}{\label{Fig:tpS_STA}}}\hspace{-2 em}
\subfigure[P$2$P Scenario: STA $0.8$ Mbps, AP $0.8$ Mbps]{\includegraphics[scale=0.52]{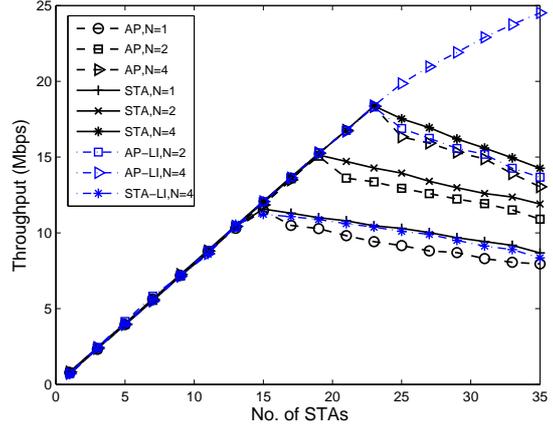}{\label{Fig:tpB_STA}}}\\
\caption{Throughput against $M$}\label{Fig:tp_STA}
\end{center}
\end{figure*}

Figure \ref{Fig:tpS_STA} shows the throughput by increasing the number of STAs in the downlink-dominant traffic scenario. It is with clear advantage to employ a higher number of antennas at the AP. The downlink throughput is much higher than the uplink one before the system gets saturated. The reasons for that are twofold: 1) the AP traffic load is inherently higher than that of STAs, and 2) the AP adopts the frame aggregation scheme. As the system becomes saturated, the throughput of both downlink and uplink decreases as $M$ increases.

As shown in Figure \ref{Fig:tpS_STA}, the uplink throughput of LI-MAC ($N=4$) is the same as that of Uni-MUMAC ($N=1$), which is because LI-MAC adopts the baseline DCF in the uplink. As the uplink throughput approaches saturation ($M=15$), the downlink throughput of LI-MAC starts to decrease. The downlink throughput of Uni-MUMAC can achieve higher gains when the network is not saturated, which is because the proposed $2$-nd round transmission increases the uplink transmission efficiency, and therefore decreases the number of AP's channel contenders. However, as the number of STAs further increases, where both up/down-link saturate, LI-MAC outperforms Uni-MUMAC, which is because Uni-MUMAC suffers a high collision rate in the 2-nd round that prolongs the 2-nd round duration. However, it is important to point out that neither LI-MAC or Uni-MUMAC is able to work sustainably in the saturated condition.

Figure \ref{Fig:tpB_STA} shows the throughput against $M$ in the down/up-link balanced traffic scenario. As expected, Uni-MUMAC achieves the balanced downlink and uplink throughput. This is because the AP and STAs are set to have the same traffic load, and more importantly, the frame aggregation scheme (AP's $N_{\text{f}} \leq M$, STA's $N_{\text{f}} =1$) counteracts the STAs' collective advantage on the channel access.

Comparing with Uni-MUMAC, the downlink throughput of LI-MAC achieves better performance when the uplink is saturated, which is because the duration of collisions in the uplink of LI-MAC is much shorter than that of Uni-MUMAC. However, the drawback is that LI-MAC has a big throughput gap between the AP and STAs, which does not satisfy the traffic requirements of the considered scenario.

\begin{figure*}[h!!!!!!]
\begin{center}
\subfigure[Downlink-dominant: STA $0.8$ Mbps, AP $3.2$ Mbps]{\includegraphics[scale=0.52]{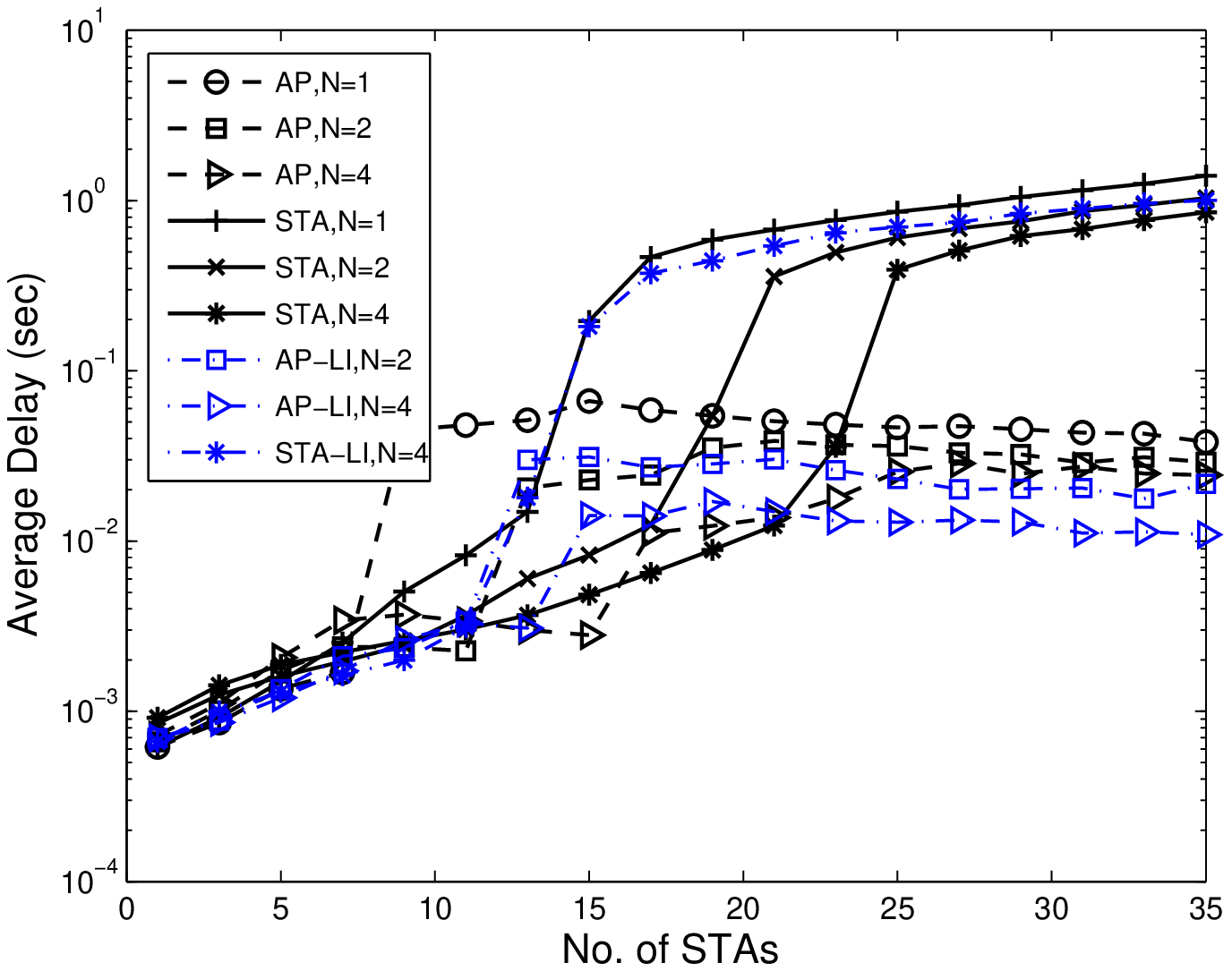}{\label{Fig:delayS_STA}}}\hspace{-2 em}
\subfigure[P$2$P Scenario: STA $0.8$ Mbps, AP $0.8$ Mbps]{\includegraphics[scale=0.52]{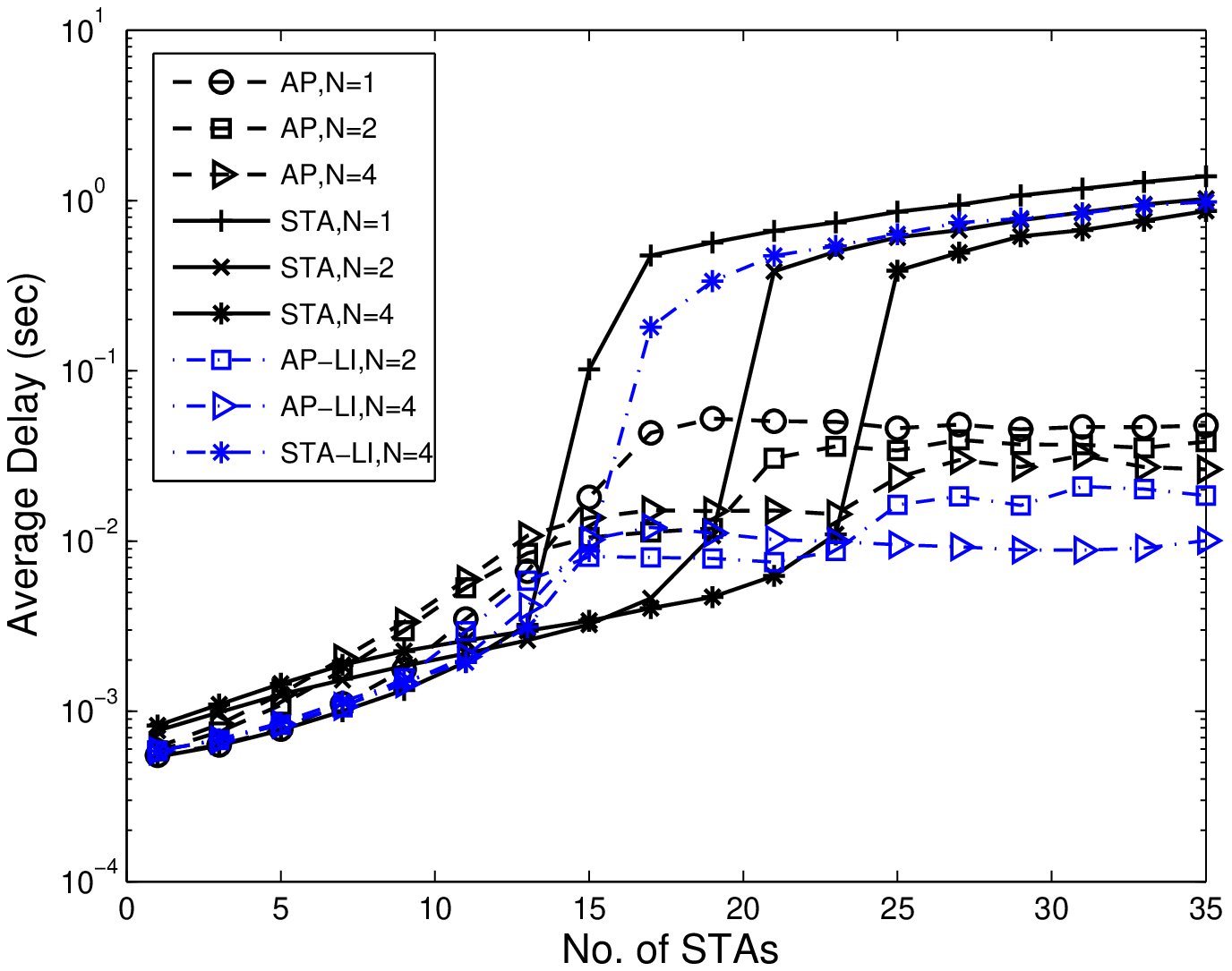}{\label{Fig:delayB_STA}}}\\
\caption{Average delay against $M$}\label{Fig:delay_STA} 
\end{center}
\end{figure*}

Figure \ref{Fig:delay_STA} shows the average delay against $M$. Both downlink and uplink delays increase with $M$, and grow significantly as the downlink or the uplink traffic approaches the saturation. After the system gets saturated, the average delay becomes steady. It is worth pointing out that the average delay of STAs is higher than that of the AP when $M$ becomes bigger. The reason for that is that the transmission duration of the AP gets longer as $M$ increases (due to the frame aggregation scheme), which makes STAs waiting longer to access the channel. 

Figure \ref{Fig:cp_STA} shows the $1$-st round collision probability increases with $M$ and converges when the system becomes saturated, which confirms the down/up-link saturation trend as discussed in Figures \ref{Fig:tp_STA} and \ref{Fig:delay_STA}. It is interesting to note that the collision probability of STAs is higher than that of the AP when the system is non-saturated. The reason for that is a STA transmits less frequently than the AP in the non-saturated condition, which results in a lower conditional collision probability for the AP. It can be clearly explained by Equation \ref{eq:1stCP}, where $p_\text{ap}$ and $\tau_\text{ap}$ ($p_\text{sta}$ and $\tau_\text{sta}$) are the $1$-st round collision probability and the transmission probability of the AP (or a STA) in the non-saturated condition:

\begin{figure*}[t!!!!!!]
\begin{center}
\subfigure[Downlink-dominant: STA $0.8$ Mbps, AP $3.2$ Mbps]{\includegraphics[scale=0.52]{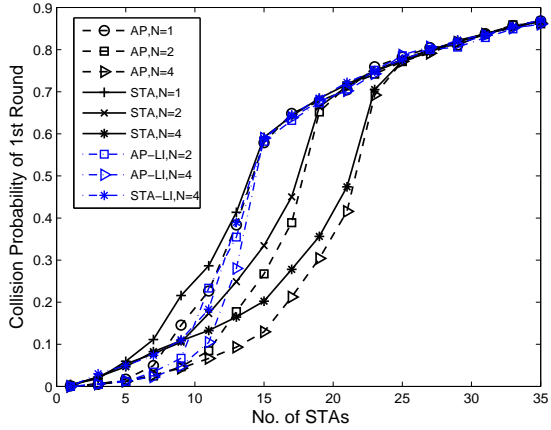}{\label{Fig:cpS_STA}}}\hspace{-2 em}
\subfigure[P$2$P Scenario: STA $0.8$ Mbps, AP $0.8$ Mbps]{\includegraphics[scale=0.52]{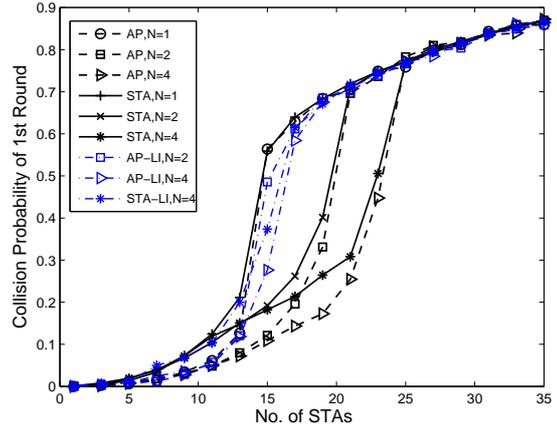}{\label{Fig:cpB_STA}}}\\
\caption{$1$-st round collision probability against $M$}\label{Fig:cp_STA}
\end{center}
\end{figure*}
\begin{equation}
\begin{aligned}
\begin{cases}
  p_\text{ap}=1-(1-\tau_\text{sta})^{M} \\
  p_\text{sta}=1-(1-\tau_\text{sta})^{M-1}\cdot(1-\tau_\text{ap}).
\label{eq:1stCP}
\end{cases}
\end{aligned}
\end{equation}

\begin{figure}[h!!!!!!!!]
\centering
\includegraphics[scale=0.55]{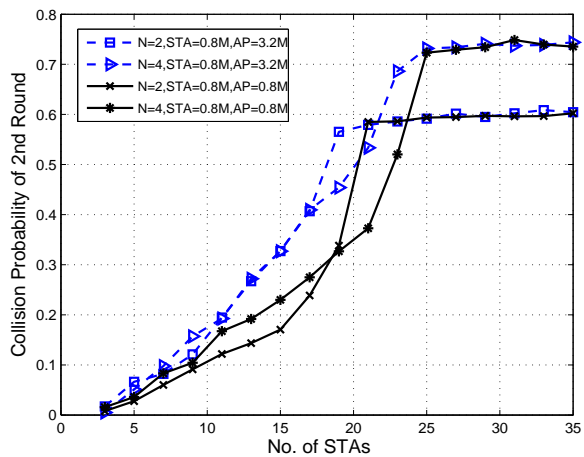}
\caption{$2$-nd round collision probability against $M$}
\label{Fig:2ndcp_STA}
\end{figure}

Figure \ref{Fig:2ndcp_STA} shows the $2$-nd round collision probability against $M$. It is clear that the $2$-nd round collision probability is higher when the system traffic load is higher. In the low number of STAs area, the $2$-nd round collision probability when the AP has $2$ antennas is sometimes lower than that when the AP has $4$ antennas. The reason is that, a higher number of antennas at the AP usually means a longer duration of the $2$-nd contention round, which increases the chances of collisions in the $2$-nd round. For example, in a case that the AP employs $2$ antennas, the $2$-nd contention round finishes as soon as a STA successfully wins the still-available antenna of the AP; while in a case that the AP employs more than $2$ antennas, the $2$-nd contention round continues, therefore increasing the $2$-nd round collision probability.
\section{Conclusions \& Future Research Challenges} \label{sec:conclusions}
In this paper, a unified MU-MIMO MAC protocol called Uni-MUMAC, which supports both MU-MIMO downlink and uplink transmissions for IEEE 802.11ac WLANs, is proposed. We evaluate it through an analytic model and simulations. A prominent MAC scheme from the literature is implemented and compared with Uni-MUMAC. 

By analyzing the simulation results, we observe that the $2$-nd round Contention Window $CW_{\text{2nd}}$, which is tuned to optimize the uplink transmission, is however not bringing the same benefit to the downlink one. An adaptive frame aggregation scheme and a queue scheme are applied at the AP to offset this disadvantage. By properly setting the aforementioned parameters, the results show that a WLAN implementing Uni-MUMAC is able to avoid the AP bottle-neck problem and performs very well in both the traditional downlink-dominant and emerging down/up-link balanced traffic scenarios. The results also show that a higher system capacity can be achieved by employing more antennas at the AP.

Uni-MUMAC gives us insight about the interaction of down/up-link transmissions and how different parameters that control the system can be tuned to achieve the maximum performance. Based on the study of this paper, we considered the following aspects as the future research challenges or next steps for Uni-MUMAC.
\begin{enumerate}
\item{\textbf{Adaptive Scheduling Scheme}}: As discussed in the paper, a parameter that optimizes the uplink could be unfavorable to the downlink. Therefore, an adaptive scheduling algorithm that takes several key parameters into account and compensates those STAs whose interests are harmed would play a significant role on obtaining the maximum performance while maintaining the fairness. As implied from the results, these parameters include: the size of A-MPDU, the queue length, the spatial-stream/frame allocation, the number of nodes/antennas, and other key parameters that control down/up-link transmissions.
\item{\textbf{Traffic Differentiation}}: Another future research challenge is to provide new traffic differentiation capability in the uplink in addition to the one defined in IEEE 802.11e amendment \cite{11e}. The new traffic differentiation should be able to limit the number of STAs that can participate in the $2$-nd contention round to reduce $2$-nd round collisions. A possible solution could be to create a table at the AP with information about the priority of each traffic flow and the queue length of each STA, and then to utilize this table to control the $2$-nd contention round.
\item{\textbf{Multi-hop Mesh Networks}}: There are more challenges that need to be considered in designing MAC to operate in multi-hop wireless networks. First, the hidden-node problem. It is still an open challenge to find mechanisms that efficiently solve the collisions caused by hidden nodes. A collision-free scheme proposed in \cite{barcelo2012distributed} or the handshake based coordinated access could be a starting point to combat the hidden-node collisions in wireless mesh networks. Secondly, due to the heterogeneity of mesh nodes (e.g., different number of antennas at nodes), MAC protocols for wireless mesh networks need to be designed with the capability of swiftly switching among MU-MIMO, SU-MIMO, multi-packet and single-packet transmission schemes. Thirdly, MAC and routing protocols need to be jointly designed. There could be multiple destinations involved in a MU-MIMO transmission, and some destinations could be out of the one-hop transmitting range, in which case, routing strategies should be able to forward multiple packets to different nodes in parallel.
\end{enumerate}
\begin{acknowledgements}
This work has been supported by the Spanish Government and the Catalan Government under projects TEC2012-32354 (Plan Nacional I+D), CSD2008-00010 (Consolider-Ingenio Program) and SGR2009\#00617.
\end{acknowledgements}

\bibliographystyle{ieeetr}

%
%
\bibliography{UpDownRefs}

\end{document}